\documentclass[aps,twocolumn,showpacs,amsmath,amssymb,pra,superscriptaddress,floatfix,longbibliography]{revtex4-2}
\usepackage{lipsum}
\usepackage{hyperref}
\usepackage{bm}% bold math
\usepackage{physics}
\usepackage{braket}
\usepackage{amsmath}
\usepackage{amssymb}
\usepackage{mathtools}
\usepackage{graphicx}% Include figure files
\usepackage{dcolumn}% Align table columns on decimal point
\usepackage{bm}% bold math
\usepackage{multirow}
\usepackage{comment}
\usepackage{leftidx}
\usepackage{subcaption} % DO NOT USE! INTERFERES WITH REVTEX!!!

\usepackage[normalem]{ulem}
\usepackage{color}
\usepackage{xcolor}
\usepackage{float}
\usepackage{capt-of}
\usepackage{mwe}
\usepackage{amsfonts}
\usepackage[capitalise]{cleveref}
\usepackage[shortlabels]{enumitem}
\usepackage[german,english]{babel}
\usepackage{gensymb}
\usepackage[thinc]{esdiff}
\usepackage{url}
\graphicspath{ {./Images/Free}, {./Images/bound} }
\setlist[enumerate,1]{label={(\roman*)}}
\begin{document}

%\preprint{}

\title{Quantum Interference in Two-Atom Resonant X-ray Scattering of an Intense Attosecond Pulse}
\author{Akilesh Venkatesh}
\email[]{avenkatesh[at]anl[dot]gov}
\affiliation{Chemical Sciences and Engineering Division, Argonne National Laboratory, Lemont, Illinois 60439, USA}

\author{Phay J. Ho}
\email[]{pho[at]anl[dot]gov}
%\thanks{}
\affiliation{Chemical Sciences and Engineering Division, Argonne National Laboratory, Lemont, Illinois 60439, USA}


\date{\today}

\begin{abstract}
We theoretically investigate resonant x-ray scattering from two non-interacting Ne\textsuperscript{+} ions driven by an intense attosecond pulse using a non-relativistic, QED-based time-dependent framework. Our model includes Rabi oscillations, photoionization, Auger decay, and quantum interference among elastic scattering and resonance fluorescence pathways. We analyze how the total scattering signal depends on pulse intensity, atomic configuration, and initial electronic state.  We find that the total resonant scattering yield exceeds its non-resonant counterpart; the angular dependence of the signal qualitatively resembles a two-atom structure factor; and the visibility of interference fringes is sensitive to pulse area and the initial electronic state. Only a subset of final states reached via resonance fluorescence exhibits interference, determined by the indistinguishability of photon emission pathways. Fringe visibility is maximized in the linear scattering regime, where ionization is minimal and resonance fluorescence pathways can be largely indistinguishable. These results highlight optimal conditions for applying ultrafast resonant x-ray scattering to single-particle imaging.
\end{abstract}

%\keywords{}

\maketitle

%\section{Figures and Captions} \label{Section_introduction}

\section{Introduction}

The ability to extract structural information from x-ray scattering lies at the heart of modern imaging science. Non-resonant elastic scattering, first explained by Bragg over a century ago~\cite{braggdiffraction1913}, remains foundational for crystallography by linking interference patterns to periodic atomic structures. Due to its low scattering cross section, non-resonant scattering typically requires crystalline order to achieve sufficient signal. 
Beyond crystallography, coherent diffractive imaging (CDI) \cite{miao1999extending,miao2025computational} extends structure determination to isolated, non-crystalline samples by applying phase-retrieval algorithms \cite{fienup1982phase,miao1998phase,marchesini2003x} to measured x-ray diffraction patterns.
The advent of x-ray free-electron lasers (XFELs)~\cite{XFEL1,XFEL_SLAC_progressreview,LCLS_5yrs,XFEL_theory,XFEL_DESY_progressreview,EuropeanXFEL1,EuropeanXFEL2,EuropeanXFEL3,SACLA_1,SACLA_2,SwissFEL1,SwissFEL2,SwissFEL3} has further inspired ultrafast imaging of atomic-scale structures \cite{chapman2011femtosecond, seibert2011single, ekeberg2015three} and non-equilibrium dynamics \cite{Gorkhover-2016-NatPho, gomez2014shapes,ferguson2016transient}. These sources deliver intense, femtosecond-to-attosecond x-ray pulses~\cite{LCLS_specswebsite,Euxfel_specswebsite,SACLA_specs_website1,SACLA_specs_website2} enabling single-particle imaging (SPI), a CDI-based approach in which diffraction from an individual particle 
is recorded in a single x-ray shot. SPI relies on the principle of ``imaging before destruction" \cite{neutze2000_imagebeforedestroy}, where the goal is to capture structural information before radiation-induced ionization and fragmentation degrade the sample.

Yet, even with XFELs, the low cross section of non-resonant scattering necessitates extreme intensities that often trigger rapid ionization and sample destruction~\cite{Young2010,Rudek-2012-NatPho,Doumy-PRL-2011,Hoener-PRL-2010,Schorb-PRL-2012,Bostedt2012,Ho-2016-PRA}.  To overcome this limitation, resonant scattering from transient ionic states has been explored~\cite{Hiddenresonance_kanter,Ho-2020-NatComm}, demonstrating enhanced yield and contrast in xenon clusters~\cite{Tais_2022_Xenon}.   The development of attosecond XFEL pulses\cite{attoXLEAP2020} with temporal coherence has further advanced SPI by reducing damage and at the same time motivated a study that exploits x-ray-driven Rabi dynamics in neon clusters~\cite{Tais_Ne_apstalk}.

The quantum nature of interference underlying resonant scattering has been studied in the optical domain. Richter~\cite{Richter1991_RFinterference} theoretically predicted first-order interference in resonance fluorescence from two non-interacting atoms. Eichmann, Itano and co-workers ~\cite{Eichmann_Youngs_1993, Eichmann_NISTreport_1996}, using a two-ion system as an analog of Young's double-slit experiment, conclusively demonstrated the interference pattern in resonantly scattered light under weak-field, monochromatic conditions. These optical studies showed that the phenomenon could be understood through “which-path’’ arguments~\cite{itano1998_Youngs}; that is, the indistinguishability of the resonance fluorescence pathways that lead to the same final electronic state allows for interference. This requirement of state indistinguishability is uniquely quantum: pathways leading to different final electronic states are distinguishable and therefore do not interfere. This is unlike the interference in Thomson scattering which can be described classically.
A subsequent study by Agarwal et al.~\cite{agarwal2002_Iorder_vs_IIorder} further analyzed the properties of first-order (one-photon) interferences and second-order intensity-intensity (two-photon)correlations in two-particle systems. Recently, Young's double slit type interference has been reported~\cite{YDS_photoelectron_2020} in the photoelectron yield emitted from homonuclear diatomic molecules and has been used to extract the birth-time delay between electron emissions from the two atomic centers.

\begin{figure*}
%\centering
\begin{minipage}[t][3ex][t]{0.001\textwidth}
%(a)
\end{minipage}
\begin{minipage}[t]{0.95\textwidth} \vspace{0.05cm}
\includegraphics[width=\textwidth]{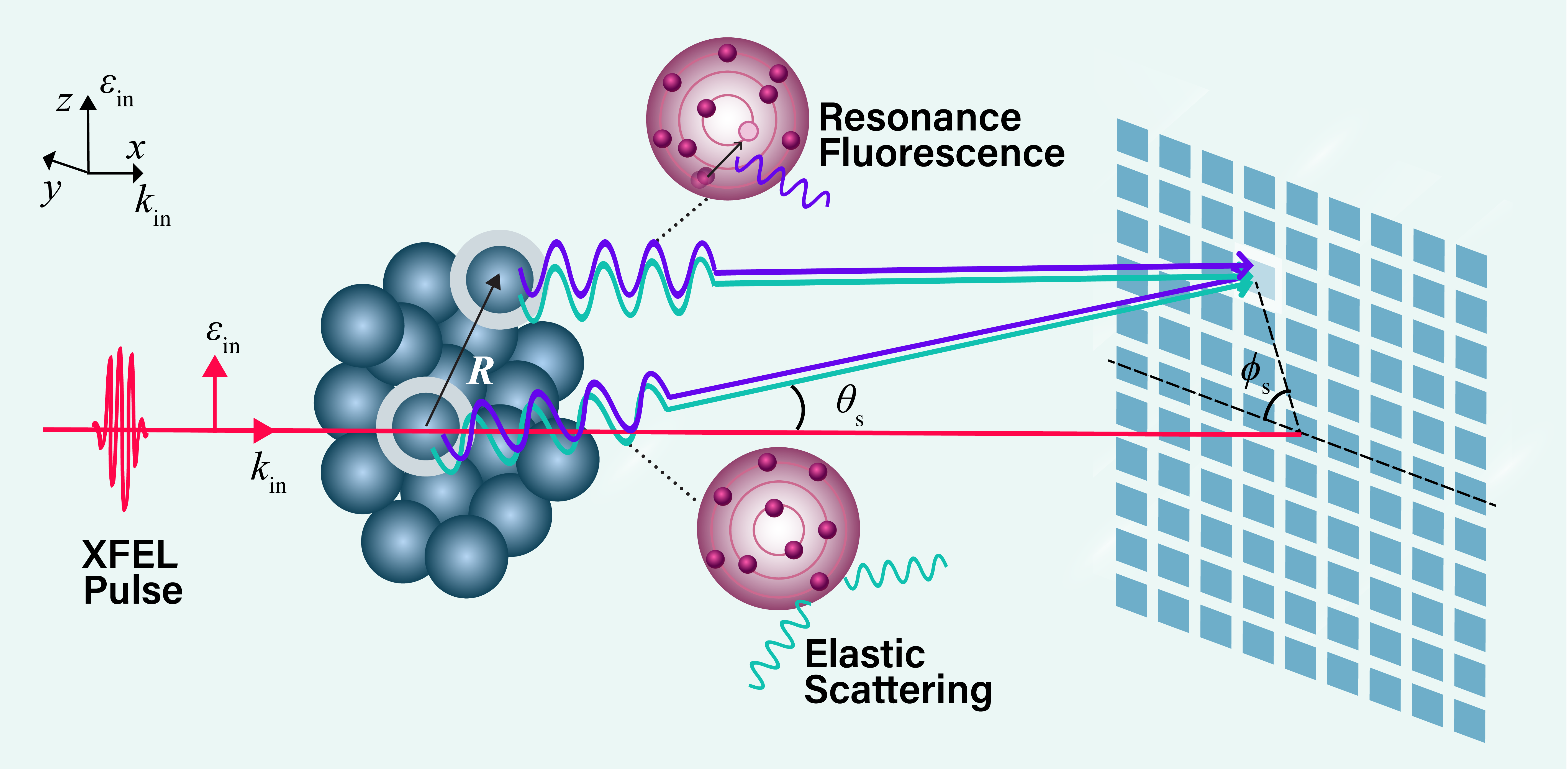}
\hfill
\end{minipage}
\caption{\label{Schematic_diagram} Schematic diagram of resonant scattering from a collection of atoms including both resonance fluorescence and elastic scattering pathways for scattering. The incident x-ray is chosen to be linearly polarized along the z-axis and propagating along the x-axis.  We note that only two atoms are included in our quantum-mechanical treatment, and the additional atoms are shown only as visual context representing a typical experimental environment.}
\end{figure*}

In the x-ray regime, one-photon interference effects have been theoretically explored under weak-field, monochromatic conditions using perturbative treatments~\cite{Gelmukhanov_1975interference_YmaRef1,Yanjun_ma_1995interference,Yanjun_ma_1994_solids}. Similar interference phenomena have also attracted interest in coherent nuclear resonance scattering~\cite{Nuclear_resonant_smirnov1995, Nuclear_resonant_kohn1998, smirnov1999_nuclearresonant, andreeva2005nuclear} with synchrotron sources. Two-photon interference approaches based on intensity correlations~\cite{HBT1956correlation,Richter1991_RFinterference, Ho-2021-SD, Ho-2023-Physics} were proposed, first by Classen and co-workers \cite{Classen-2017-PRL}, as a means to extract high-resolution structural information from x-ray fluorescence signals.  This method, which is also sometimes referred to as incoherent diffractive imaging \cite{Classen-2017-PRL,lohse2021incoherent}, has recently been demonstrated experimentally to image x-ray fluorescing structures in copper ~\cite{Trost-2023-PRL} and vanadium foils \cite{radloff2025fluorescence}. However, questions remain regarding how strong-field ultrafast dynamics, such as Rabi oscillations, photoionization, and Auger decay, modify first-order interference and scattering signatures. In particular, under intense x-ray pulse excitation, both elastic scattering (ES) and resonance fluorescence (RF) pathways can contribute (see Fig.~\ref{Schematic_diagram}), and their interference depends critically on the indistinguishability of the final atomic states.

In prior work~\cite{Res_Rabi_PRAL,Res_theory_PRA}, we developed a QED-based time-dependent Schr\"odinger equation (TDSE) framework to model resonant scattering from a single Ne\textsuperscript{+} ion under intense ultrafast x-ray pulses. Here, we extend this approach to study coherent scattering from two non-interacting Ne\textsuperscript{+} ions, incorporating Rabi dynamics, photoionization, Auger decay, ES, and RF. We analyze how the total scattered signal depends on interatomic separation, geometry, pulse parameters, and initial states. Interference between atoms is shown to encode structural information in the angular distribution, and the total scattered yield under resonant conditions approaches a structure factor form akin to non-resonant diffraction. Fringe visibility is maximum in the linear x-ray scattering regime, where fluorescence pathways can interfere coherently and ionization is small. These results offer insights about the optimum parameters for using ultrafast resonant scattering in structural imaging.

This paper is organized as follows: We first describe an approach for studying resonant ultrafast x-ray scattering response of two non-interacting atoms in Sec. \ref{Methods}. We use the approach to investigate the two-atom scattering response in Sec.~\ref{Sec_results}. We present a summary and outlook in Sec. \ref{conclusion_summary}.  Unless otherwise stated, atomic units are used in the equations presented in this work.

\section{Methods and modelling} \label{Methods}
We employ a QED-based time-dependent scattering theory approach~\cite{NLCPRA_1} that was previously used to study the single-atom resonant response \cite{Res_theory_PRA}. This approach captures processes that involve multiple incident photons but one outgoing photon. A detailed derivation of the approach can be found in Ref.\cite{Res_theory_PRA}. In this section, we extend that approach to study the scattering response from two non-interacting atoms of the same kind.
%\subsection{Bound state contribution for scattering amplitude} \label{Subsec_Bound_state_contrb}

The total vector potential $\boldsymbol{\hat{A}}$ of the electromagnetic field is written as a sum of a classical incident field $\boldsymbol{A}_C$ and a quantized outgoing field $\boldsymbol{ \hat{A}}_{Q}$. The total wavefunction ansatz is,
\begin{equation} \label{wfn_ansatz_Nelectron}
    \ket{\psi_{total}(t) } = \psi^{(0)} (t) \ket{0} 
    + \sum_{\boldsymbol{k},\boldsymbol{\epsilon}} \psi_{\boldsymbol{k},\boldsymbol{\epsilon}} ^{(1)}(t) e^{-i \omega_{k} t} {\hat{a}_{\boldsymbol{k},\boldsymbol{\epsilon}} }^{\dagger}  \ket{0}.
\end{equation}
Here $\psi^{(0)}$ describes the unscattered wave at the time $t$ and it captures the interaction of the system with the incident classical field. In other words, the unscattered wave describes the electronic state if there is no scattering. The quantity $\psi_{\boldsymbol{k},\boldsymbol{\epsilon}} ^{(1)}(t)$ describes the state of the scattered system and the scattering amplitude for a photon to scatter with momentum $\boldsymbol{k}$ and polarization $\boldsymbol{\epsilon}$.

The incident x-ray field is chosen to be a linearly polarized Gaussian pulse and is described as,
\begin{equation}\label{classicalvectorpotential}
\begin{split} 
\boldsymbol{A}_C (\boldsymbol{r},t) = A_0(\boldsymbol{r}, t) \sin\big(  {\boldsymbol{k}_{in} \cdot \boldsymbol{r} } - \omega_{in} t \big)    \boldsymbol{\epsilon}_{in} ,
\end{split}
\end{equation}
where $\boldsymbol{r}$ refers to the position vector. $\boldsymbol{k}_{in}$, $\omega_{in}$ and $\boldsymbol{\epsilon}_{in}$ are the momentum, energy and polarization of the incident photons, respectively. Also, $\omega_{in} = c\abs{\boldsymbol{k}_{in}} $, where $c$ is the speed of light in vacuum. The envelope for the Gaussian pulse is,
\begin{equation}\label{A0_gaussian}
A_0(\boldsymbol{r}, t) = \frac{E_{in}}{\omega_{in}} \exp\Bigg[\frac{(-(2 \ln{2} ) (t - \frac{{\boldsymbol{\hat{k}}_{in} \cdot \boldsymbol{r} }}{c})^2)}{t^2_{wid}} \Bigg],
\end{equation}
where $E_{in}$, and $t_{wid}$ are the incident electric field amplitude and pulse duration (full width half-maximum of the intensity), respectively. The quantized scattered field $\boldsymbol{ \hat{A}}_{Q}$ is given by~\cite{Loudon},
\begin{equation}\label{APDX_Eqn_quantized vector potential}
    \boldsymbol{\hat{A}}_Q
    = \sqrt{ \frac{2\pi}{  V  } } \sum_{\boldsymbol{k},\boldsymbol{\epsilon}}  \frac{1}{\sqrt{\omega_{k}}} \bigg[  \boldsymbol{\epsilon} e^{i\boldsymbol{k\cdot r}} \hat{a}_{\boldsymbol{k},\boldsymbol{\epsilon}} +  \boldsymbol{\epsilon}^* e^{-i\boldsymbol{k\cdot r} }
    { \hat{a}_{\boldsymbol{k},\boldsymbol{\epsilon}} ^\dagger  }\bigg].
\end{equation}
Here, $V$ denotes the quantization volume. The operators $\hat{a}_{\boldsymbol{k},\boldsymbol{\epsilon}}^{\dagger}$ and $\hat{a}_{\boldsymbol{k},\boldsymbol{\epsilon}}$ denote the creation and annihilation operators for a photon with momentum $\boldsymbol{k}$ and polarization $\boldsymbol{\epsilon}$. Here $\boldsymbol{k} \cdot \boldsymbol{\epsilon} = 0$.

The total Hamiltonian for the system of two identical and non-interacting atoms exposed to an intense attosecond x-ray pulse is,
\begin{equation} \label{full hamiltonian}
    \hat{H}_{tot} = \sum\limits_{ j=1}^{2} \hat{H}(\boldsymbol{R_j}) + \sum_{\boldsymbol{k},\boldsymbol{\epsilon}}\omega_{k} \hat{a}_{\boldsymbol{k},\boldsymbol{\epsilon}}^{\dagger} \hat{a}_{\boldsymbol{k},\boldsymbol{\epsilon}}.
\end{equation}
Here $\boldsymbol{R}_j$ is the position of the $j$-th atom, each with $n_e$ electrons, and
\begin{equation}\label{Hamiltonian_2}
\begin{split}
  \hat{H}(\boldsymbol{R}_j) = & \sum\limits_{ b_j=1}^{n_e} \bigg[\frac{(\boldsymbol{\hat{P}}_{b_j} + \boldsymbol{\hat{A}}(\boldsymbol{R}_j + \boldsymbol{r}_{b_j}))^2}{2} + \hat{V_a}(|\boldsymbol{R}_j - \boldsymbol{r}_{b_j}|) \bigg]  \\
  &+ \sum\limits_{{i=1,b_j>i}}^{n_e} \hat{V}_{ee}(|\boldsymbol{r}_{b_j}-\boldsymbol{r}_{i}|) ,
\end{split}
\end{equation}
where $\hat{\boldsymbol{P}}_{b_j}$ and $\hat{V}_a(|\boldsymbol{R}_j - \boldsymbol{r}_{b_j}|)$ describe the quantum mechanical operator for momentum and for potential energy between the nucleus and the $b_j$\textsuperscript{th} electron of the $j$-th atom, respectively. The quantity $\hat{V}_{ee}$ describes electron-electron repulsion. We neglect the Coulombic interaction and the associated nuclear motion between the two atoms during this scattering process as the time-scale of the nuclear motion is much larger than the pulse duration and the Auger-lifetime of the atoms.

Using the time-dependent Schr\"odinger equation (TDSE) for the total wavefunction [Eq.~(\ref{wfn_ansatz_Nelectron})] and the total Hamiltonian [Eq.~(\ref{full hamiltonian})] and following a similar approach to Ref.~\cite{Res_theory_PRA}, we obtain equations for $\psi^{(0)}$ and $\psi_{\boldsymbol{k},\boldsymbol{\epsilon}} ^{(1)}(t)$,
\begin{widetext}
\begin{equation} \label{TDSE_psi0}
  i \frac{\partial \psi ^ {(0)} }{\partial t} - \sum\limits_{j=1}^{2}\sum\limits_{ b_j=1}^{n_e} \hat{H}_{C} (\boldsymbol{R}_j + \boldsymbol{r}_{b_j}, t) \psi ^{ (0) } = 0,
\end{equation}
\begin{equation} \label{Exact_eqn_psi1}
\begin{split} 
  i \frac{\partial \psi ^ {(1)}_{\boldsymbol{k},\boldsymbol{\epsilon}} }{\partial t} - \sum\limits_{j=1}^{2}\sum\limits_{ b_j=1}^{n_e} \hat{H}_{C} (\boldsymbol{R}_j + \boldsymbol{r}_{b_j}, t) \psi ^{ (1) }_{\boldsymbol{k},\boldsymbol{\epsilon}} =  & \sqrt{\frac{2\pi}{ V\omega_{k}} } e^{i\omega_{k} t }  \sum\limits_{j=1}^{2}\sum\limits_{b_j=1}^{n_e} e^{-i\boldsymbol{k}\cdot (\boldsymbol{R}_j + \boldsymbol{r}_{b_j})} ~
  \boldsymbol{\epsilon}^* \cdot \bigg[\boldsymbol{\hat{P}}_{b_j} + \boldsymbol{A}_{C}(\boldsymbol{R}_j +  \boldsymbol{r}_{b_j}, t) \bigg] W(t) \psi ^{(0)},
\end{split}
\end{equation}
\end{widetext}
where $V$ is the quantization volume, $W(t)$ is a windowing function that turns on and off adiabatically the interaction with the quantized field, and 
\begin{equation} \label{definition_H_C}
\begin{split}
    \hat{H}_{C} (\boldsymbol{r}_b, t) = & \frac{(\boldsymbol{\hat{P}}_b + \boldsymbol{A}_{C} (\boldsymbol{r}_b, t))^2}{2} + \hat{V}_{a}(|\boldsymbol{r}_b|)  \\
    & + \sum\limits_{{i=1,b>i}}^{n_e} \hat{V}_{ee}(|\boldsymbol{r}_b-\boldsymbol{r}_i|) ~.        
\end{split}    
\end{equation}
The above equations [Eqs.~(\ref{TDSE_psi0}) \& (\ref{Exact_eqn_psi1})] were derived by treating the incident field non-perturbatively and the scattered field perturbatively.

The combined state of the two non-interacting atoms can be expanded in a basis of tensor product of single-atom eigenstates. Therefore, the expansion of the $\psi^{(0)}$ and $\psi_{\boldsymbol{k},\boldsymbol{\epsilon}} ^{(1)}(t)$ take the following form,
\begin{equation} \label{Expand_psi0_eigenstates}
\ket{\psi^{(0)} (t)} = \sum\limits_{m_1,m_2}^{n_s} C^{(0)}_{m_1 m_2}(t)  \ket{\psi_{m_1},\psi_{m_2}},
\end{equation}
\begin{equation} \label{Expand_psi1_eigenstates}
\ket{\psi_{\boldsymbol{k},\boldsymbol{\epsilon}} ^{(1)}(t)} = \sum\limits_{m_1,m_2}^{n_s} C^{(1)}_{m_1 m_2;\boldsymbol{k} \boldsymbol{\epsilon}}(t)   \ket{\psi_{m_1},\psi_{m_2}}.
\end{equation}
$C^{(0)}_{m_1m_2}$ describes the probability amplitude for the two atoms located at $\boldsymbol{R}_1, \boldsymbol{R}_2$ to be in states $\ket{\psi_{m_1}}$ and $\ket{\psi_{m_2}}$, respectively. Here $n_s$ is the number of bound states included for a single atom. Also note that $|C^{(0)}_{m_1m_2}|^2$ effectively describes the population for the state $\ket{\psi_{m_1},\psi_{m_2}}$.
Similarly $C^{(1)}_{m_1 m_2;\boldsymbol{k} \boldsymbol{\epsilon}}(t)$ describes the entangled probability amplitude for the scattered photon to have momentum $\boldsymbol{k}$ and polarization $\boldsymbol{\epsilon}$ \textit{and} for the two-atom system to be in the state $\ket{\psi_{m_1},\psi_{m_2}}$.

We proceed similar to the derivation from the single-atom response~\cite{Res_theory_PRA} using the the expansion from Eqs.~(\ref{Expand_psi0_eigenstates}) and (\ref{Expand_psi1_eigenstates}) in Eqs.~(\ref{TDSE_psi0}) and (\ref{Exact_eqn_psi1}). 
In computing ($\psi^{(0)}$), Auger decay and photoionization are included through non-Hermitian decay terms, which lead to population loss in $|\psi^{(0)}|^2$.  Since our interest lies in the total scattered photon yield, these decay channels are omitted in the evolution of the scattered state. While including them would capture the subsequent decay of the final electronic states, it does not change the emitted photon yield~\cite{Res_theory_PRA}.
The equation for $C^{(0)}_{u_1 u_2}(t)$ is,
\begin{widetext}
\begin{equation} \label{eqn_psi0_nstate}
i \frac{\partial C^{(0)}_{u_1 u_2} }{\partial t}  - C^{(0)}_{u_1 u_2}  (\xi_{u_1} + \xi_{u_2} ) 
   -  \sum\limits_{m_1,m_2}^{n_s} C^{(0)}_{m_1 m_2} \sum\limits_{j=1}^{2}  
   \boldsymbol{A}_C(\boldsymbol{R}_j,t) \cdot\bra{ \psi_{u_j} }\sum\limits_{b_j=1}^{n_e} \hat{\boldsymbol{P}}_{b_j} \ket{\psi_{m_j} } \prod_{l\neq j} \delta_{u_l m_l} = 0.
\end{equation}
\end{widetext}
Here $\delta_{u_l m_l} $ is the Kronecker delta function and $\xi_{u_j} = E_{u_j} - \frac{i}{2} \big( \gamma_{{u_j},tot}^{A} + \gamma_{{u_j},tot}^{P}(t) \big)$. The quantities $E_{u_j}$ denotes the binding energy of the state $\ket{\psi_{u_j}}$, while $\gamma_{{u_j},tot}^{A}$ and $\gamma_{{u_j},tot}^{P}(t)$ are the total Auger and the instantaneous photoionization rate from $\ket{\psi_{u_j}}$ to all Ne$^{2+}$ configurations. The time-dependent photoionization rate is given by $\gamma_{u_j,tot}^{P}(t) = \frac{I(t) \sigma_{u_j,tot}^{P} }{\omega_{in}}$ with $\sigma_{u_j,tot}^{P}$ being the associated one-photon photoionization cross section. To describe the Rabi dynamics in the atoms, the spatial variation of $\boldsymbol{A}_C$ is neglected over the size of a single atom~\cite{Res_theory_PRA} consistent with the dipole approximation. However the field experienced by the two separate atoms can be different depending on the choice of $\boldsymbol{R}_j$ and wavelength, therefore the dependence on $\boldsymbol{R}_j$ may not be neglected. Importantly, the phase difference between the incident field arriving at the two atom locations is captured in the $\boldsymbol{R}_j$ dependence.

The equation for the coefficients $C^{(1)}_{ u_1 u_2;\boldsymbol{k} \boldsymbol{\epsilon}}(t)$ in Eq.~(\ref{Expand_psi1_eigenstates}) is given by,
\begin{widetext}
\begin{equation} \label{eqn_psi1_nstate}
\begin{split}
i & \frac{\partial C^{(1)}_{u_1 u_2;\boldsymbol{k} \boldsymbol{\epsilon}} }{\partial t}  -  (E_{u_1} + E_{u_2} ) C^{(1)}_{u_1 u_2;\boldsymbol{k} \boldsymbol{\epsilon}}(t)
   -  \sum\limits_{m_1, m_2}^{} C^{(1)}_{m_1 m_2;\boldsymbol{k} \boldsymbol{\epsilon}}(t) \sum\limits_{j=1}^{2} \boldsymbol{A}_C(\boldsymbol{R}_j,t) \cdot \bra{ \psi_{u_j} }\sum\limits_{ b_j=1}^{n_e} \hat{\boldsymbol{P}}_{b_j} \ket{\psi_{m_j} } \prod_{l\neq j} \delta_{u_l m_l} \\
 & =  \sqrt{\frac{2\pi}{ V\omega_{k}} } e^{i\omega_{k} t } \sum\limits_{m_1, m_2} C^{(0)}_{m_1 m_2}(t)
 \Bigg[ \sum\limits_{j=1}^{2} e^{-i \boldsymbol{k} \cdot \boldsymbol{R}_j}\boldsymbol{\epsilon}^* \cdot \bra{ \psi_{u_j} }\sum\limits_{ b_j=1}^{n_e} \hat{\boldsymbol{P}}_{b_j} \ket{\psi_{m_j} } \prod_{l\neq j} \delta_{u_l m_l} \\
   & ~~~ - \frac{i}{2} e^{-i\omega_{in} t } ~\boldsymbol{\epsilon}^* \cdot \boldsymbol{\epsilon}_{in}   \sum\limits_{j=1}^{2}  A_0(\boldsymbol{R}_j,t) e^{i\boldsymbol{q} \cdot \boldsymbol{R}_j}
\bra{ \psi_{u_j} }\sum\limits_{ b_j=1}^{n_e} e^{i \boldsymbol{q} \cdot \boldsymbol{r}_{b_j}}  \ket{\psi_{m_j} } \prod_{l\neq j} \delta_{u_l m_l}  \Bigg] W(t).
\end{split}
\end{equation}
\end{widetext}
Here $\boldsymbol{q} = \boldsymbol{k}_{in} - \boldsymbol{k}$. It is evident from the source terms i.e. the non-homogeneous terms on the right-hand side of Eq.~(\ref{eqn_psi1_nstate}), that the first term describe the resonance fluorescence contribution and the last term describe elastic Thomson scattering contribution from all the atoms. For the elastic processes explored in this work, $\omega_k \approx \omega_{in}$ and only those terms from $\bra{ \psi_{u_j} }\sum\limits_{ b_j=1}^{n_e} e^{i \boldsymbol{q}\cdot \boldsymbol{r}_{b_j}}  \ket{\psi_{m_j} }$ with $u_j=m_j$ contribute. This quantity is the elastic scattering atomic form factor $f_{m_j}(\boldsymbol{q})$ associated with state $m_j$. We note that stimulated emission is not included in the scattering probability amplitude calculations. However, stimulated emission produces photons in the same mode of the incident field and can be obtained directly from $\psi^{(0)}$~\cite{Res_theory_PRA}.

The coupled equations for $C^{(0)}_{u_1 u_2}$ and $C^{(1)}_{u_1 u_2;\boldsymbol{k} \boldsymbol{\epsilon}}$ are solved and $C^{(1)}_{u_1 u_2;\boldsymbol{k} \boldsymbol{\epsilon}}$ is used for computing the scattering probabilities and differential cross sections~\cite{Res_theory_PRA}. The scattering probability to scatter a photon into angles $\theta_s$, $\phi_s$ and polarization $\boldsymbol{\epsilon}$ and for the final scattered state to be $\ket{\psi_{m_1}, \psi_{m_2}}$ is $P_{{m_1 m_2}; \boldsymbol{k},\boldsymbol{\epsilon}} = \abs{{ C^{(1)}_{{m_1 m_2};\boldsymbol{k} \boldsymbol{\epsilon}}} }^2$. If the final scattered state is not measured, then the scattering probability to scatter a photon with momentum $\boldsymbol{k}$ and polarization $\boldsymbol{\epsilon}$ is,
\begin{equation} \label{summed_scatteringprobability}
     P_ {\boldsymbol{k},\boldsymbol{\epsilon}} = \sum\limits_{m_1,m_2}^{} \Big| { C^{(1)}_{m_1 m_2;\boldsymbol{k} \boldsymbol{\epsilon}}}\Big|^2.
\end{equation}
Here, $\boldsymbol{k} = k(\cos\theta_s,~\sin\theta_s\cos \phi_s,~\sin \theta_s \sin \phi_s)$ and two allowed choices for outgoing photon polarization are,
\begin{equation} 
\begin{split} \label{polarization_choices}
     &\boldsymbol{\epsilon}_1 = (-\sin\theta_s,~\cos\theta_s\cos\phi_s,~ \cos\theta_s\sin\phi_s), \\
     &\boldsymbol{\epsilon}_2 = (0,~ -\sin\phi_s,~ \cos \phi_s).
\end{split}
\end{equation}
The differential cross section for the two-atom system to scatter a photon into a solid angle $d\Omega$ if its polarization is not measured is given by,
\begin{equation} \label{dcs}
        \dv{\sigma}{\Omega} =
        \frac{V \omega_{in} }{(2\pi)^3}\frac{\int \sum\limits_{ \boldsymbol{\epsilon} } P_ {\boldsymbol{k},\boldsymbol{\epsilon}} k^2dk }{\int I(t) dt } .
\end{equation}
Here $I(t)$ is the instantaneous incident intensity. One can obtain the differential cross section associated with only ES by turning off the source terms in Eq.~(\ref{eqn_psi1_nstate}) associated with RF channel and vice versa~\cite{Res_theory_PRA}. We note that the fluence $F$ of the incident pulse is given by ${\int \! I(t) dt }/\omega_{in}$. Therefore, the photon yield per solid angle from the combined scattering response of the two-atom system is given by $F\times\dv{\sigma}{\Omega}$.

While the two-atom calculations are reported for neon, we expect the description of resonant ultrafast x-ray scattering to remain valid for atoms where the dominant decay pathway of the excited state is Auger decay rather than spontaneous emission~\cite{Res_theory_PRA}. While the expressions presented above can in principle be generalized to systems with more than two atoms in bound states, there are several notable challenges. One immediate problem is that, the probability for at least one atom to ionize increases significantly as the number of atoms increases, due to photoionization and Auger decay. When this occurs, the expression derived by only including bound state contribution fails to capture the full photon yield. For systems with more than two atoms, it becomes essential to account for scattering signals arising from configurations where one or more atoms are ionized. The required treatment is nontrivial and beyond the scope of this work.

\section{Results and discussion} \label{Sec_results}

\begin{figure}
\resizebox{85mm}{!}{\includegraphics{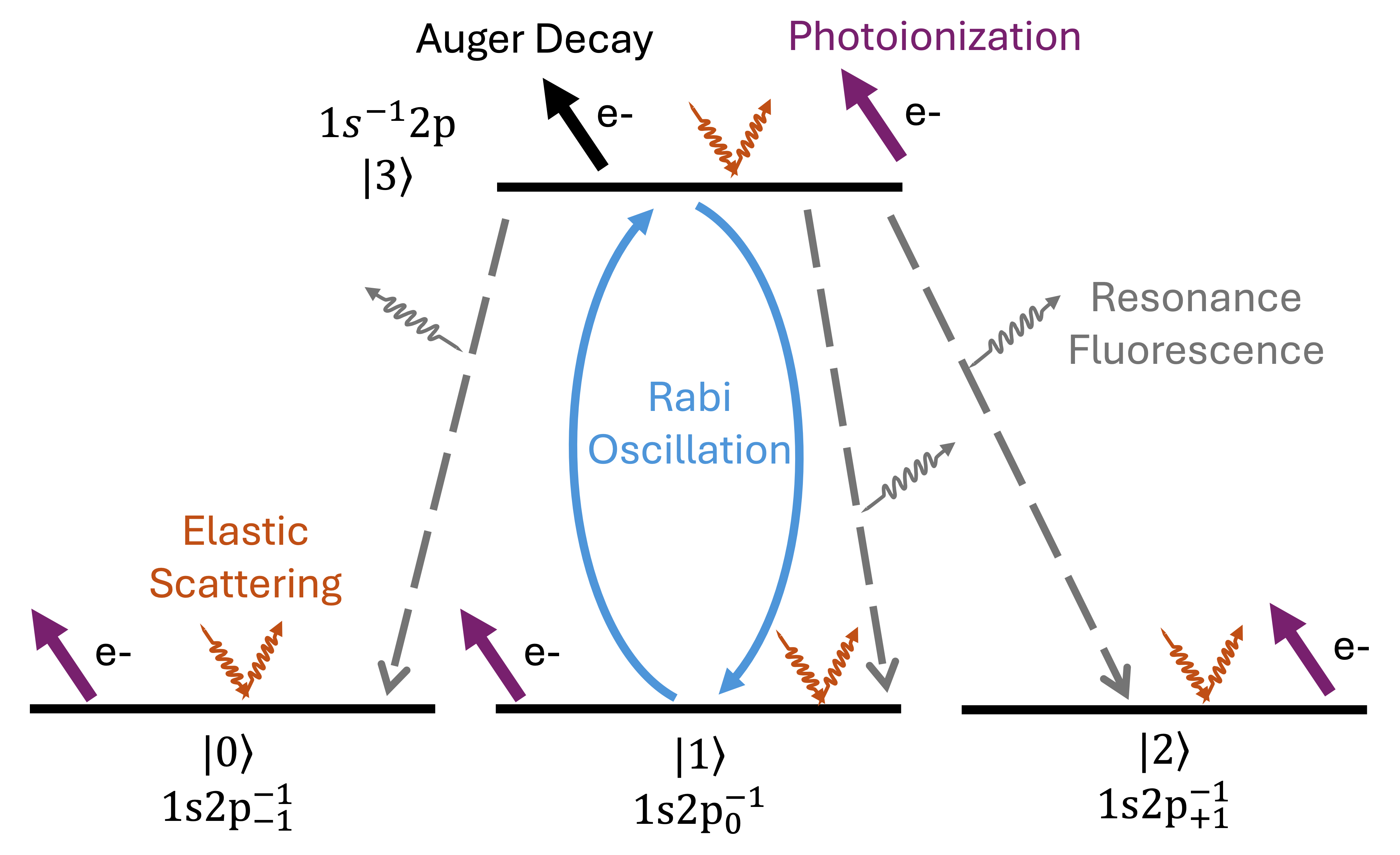}}
\caption{\label{Fig_ElectronicStates} Electronic states and processes of a single Ne\textsuperscript{+} exposed to an intense x-ray pulse. The figure includes photoionization (purple arrow) and Auger decay (black arrows) pathways and electronic transitions associated with Rabi oscillations (blue lines) and resonance fluorescence (gray dashed lines).  The elastic scattering processes (orange lines) do not induce electronic transition.}
\end{figure}

In this section, we study the resonant scattering response from two Ne\textsuperscript{+} ions, first for the case of intense pulses and then for the case of a low-intensity pulse.  The position of first atom $\boldsymbol{R}_1$ is located at the origin and the position of the second atom is displaced at $\boldsymbol{R}$ ($\boldsymbol{R}_2=\boldsymbol{R}$). Each Ne\textsuperscript{+} ion is described using the same four-level description employed in our previous single-ion study~\cite{Res_theory_PRA}.  This model include three degenerate states with a vacancy in $2p$ ($1s2p_{-1}^{-1}$, $1s2p_{0}^{-1}$ and $1s2p_{+1}^{-1}$) and a core-excited state $1s^{-1}2p$.  For our two-atom system, this results in an effective electronic basis of 16 states.  

All considered electronic processes, as shown in Fig.~\ref{Fig_ElectronicStates}, are identical to those treated in the single-ion formalism~\cite{Res_theory_PRA}.  The incident pulse is chosen (see Fig.~\ref{Schematic_diagram}) to have $\hat{\boldsymbol{k}}_{in} = \hat{x}$, $\boldsymbol{\epsilon}_{in} = \hat{z}$, a pulse duration of $t_{wid}=0.25$ fs and is resonant with the $1s2p^{-1} \rightarrow 1s^{-1}2p$ core-hole transition with a photon energy of about 849.8 eV. Depending on the incident pulse intensity it can drive Rabi oscillations between $1s2p_0^{-1}$ and $1s^{-1}2p$ in the Ne\textsuperscript{+} species, with the pulse area $Q = \int_{-\infty}^{\infty} \Omega(t) dt$. Here $\Omega(t)  = \boldsymbol{E}_{in}(t)\cdot \boldsymbol{\mu}$ is the instantaneous Rabi frequency~\cite{Cavaletto_ResFluor_PRA, Pulsearea_defn_Eberly} and $\boldsymbol{E}_{in}(t)$ and $\boldsymbol{\mu}$ refers to the instantaneous electric field amplitude and the transition dipole moment. The Ne\textsuperscript{+} ions can undergo further ionization through valence photoionization.  Also,  the core-excited state, which has a lifetime of about 2.4 fs, can undergo Auger decay. The atomic parameters for Ne\textsuperscript{+} used are identical to Ref.~\cite{Res_theory_PRA}. The dipole moment for the resonant transition $1s2p_{0}^{-1} \rightarrow 1s^{-1}2p$ is 0.0524 a.u. ~\cite{Cavaletto_ResFluor_PRA}. The photoionization cross sections were estimated to be $8.4\times 10^{-4}$ a.u. ($\sim$23.6 kilobarns) and  $1.1\times 10^{-3}$ a.u. ($\sim$31.7 kilobarns) for the degenerate ground states and core-hole state, respectively from the decay rates in Ref.~\cite{Cavaletto_ResFluor_PRA}.  The atomic form factors of Ne\textsuperscript{+} and Ne\textsuperscript{2+} are calculated using Hartree-Fock-Slater electronic structure theory \cite{Ho-2017-JPB}.

We numerically solve Eqs.~(\ref{eqn_psi0_nstate}) and (\ref{eqn_psi1_nstate}) until there is no dynamics left. This typically involves propagating the equations for much longer than the pulse and until there is no population remaining in the excited states. The actual total time the equations are solved to obtain the coherent resonant response depends on the pulse area and the positions of the atoms. For  $Q = 2\pi$ and  $Q = \pi$, the propagation times are 209 a.u. and 1588 a.u. respectively. It is worth pointing out that for a 4-level electronic state description for each atom, for a given scattering angle and outgoing photon polarization, this requires solving 32 coupled equations for a given $\omega_k$ [Eqs.~(\ref{eqn_psi0_nstate}) \& (\ref{eqn_psi1_nstate})] but for multiple $\omega_k$ points, Eq.~(\ref{eqn_psi0_nstate}) needs to be solved only once. Including the number of $\omega_k$ points required and for two outgoing photon polarizations, these translate to solving about 1984 and 3904 equations for $Q = 2\pi$ and  $Q = \pi$ pulses, respectively. These calculations are repeated for about 120 scattering angles for each angular distribution curves shown in the Secs.~\ref{subsec_initial2pz},~\ref{subsec_initial_supequal}, and~\ref{subsec_weakfield}.

\subsection{Initial state $\ket{1s2p_0^{-1}, 1s2p_0^{-1}}$} \label{subsec_initial2pz}

\begin{figure*}
\resizebox{180mm}{!}{\includegraphics {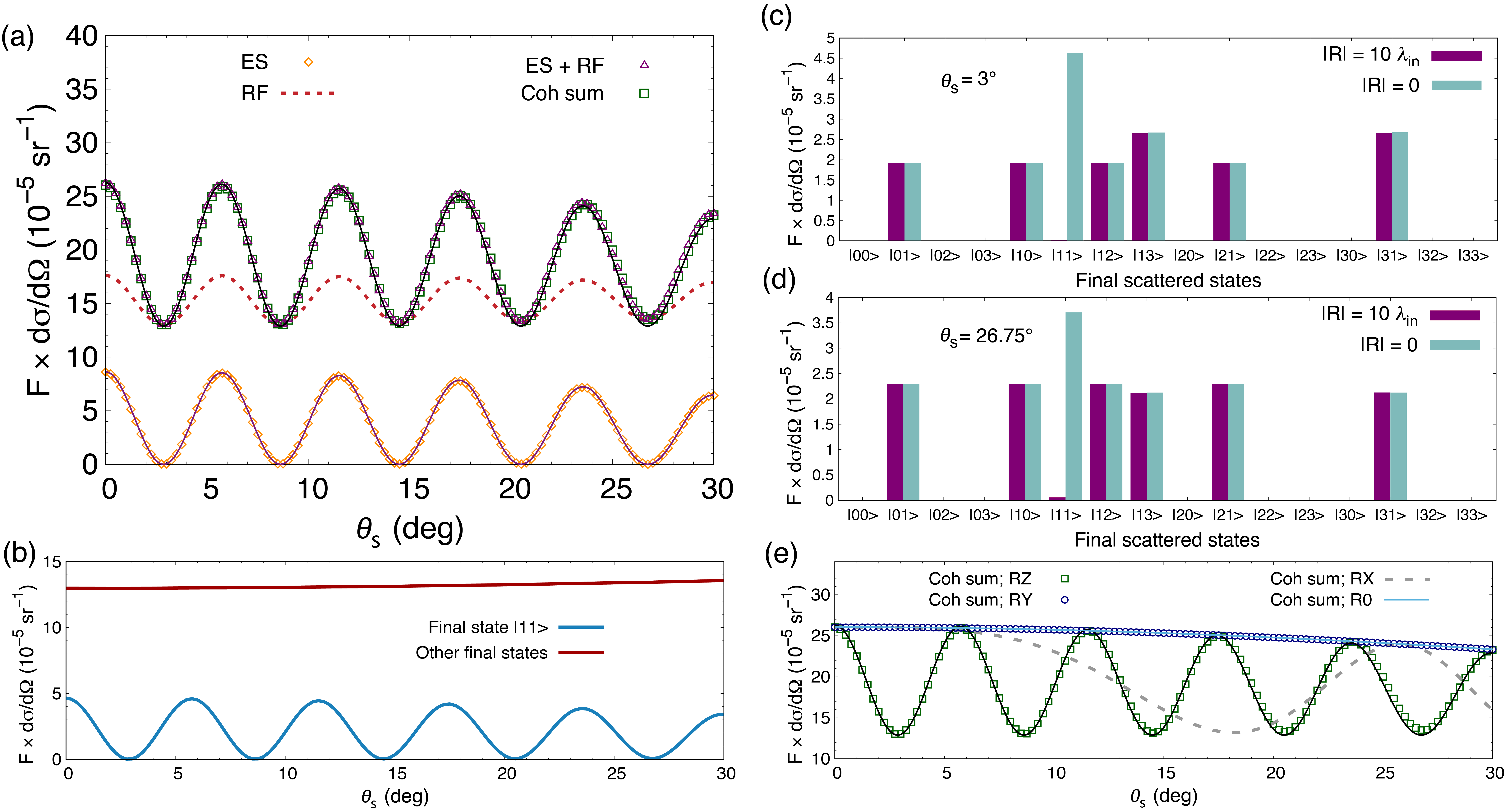}}
\caption{Resonant scattering photon yield from two atoms for a 0.25 fs ($t_{wid}$), Q = 2$\pi$ pulse for the case of planar scattering ($\phi_s = 90 \degree$) when the initial state of each atom is $\ket{1}$. (a) Scattering angle dependence for the two scattering pathways of resonance fluorescence (RF) and elastic scattering (ES), their incoherent sum (ES + RF), coherent sum (Coh sum). The black and purple solid curves are fits [Eq.~(\ref{fit_function})] which are proportional to the structure factor and $\boldsymbol{R} =10 \lambda_{in} \hat{z}$. (b) Angle dependence for resonance fluorescence contribution from different final scattered states for the same position of two atoms. (c), (d) Position dependence of resonance fluorescence  yields from different final scattered states. (e) Position dependence of the coherent sum. R0, RX, RY and RZ corresponds to $\boldsymbol{R}$=0, $\boldsymbol{R} =10 \lambda_{in} \hat{x}$, $\boldsymbol{R} =10 \lambda_{in} \hat{y}$, and $\boldsymbol{R} =10 \lambda_{in} \hat{z}$, respectively. 
}
\label{Fig_2Pi_signal_initialstate2pz}
\end{figure*}
\begin{figure*}
\resizebox{180mm}{!}
{\includegraphics {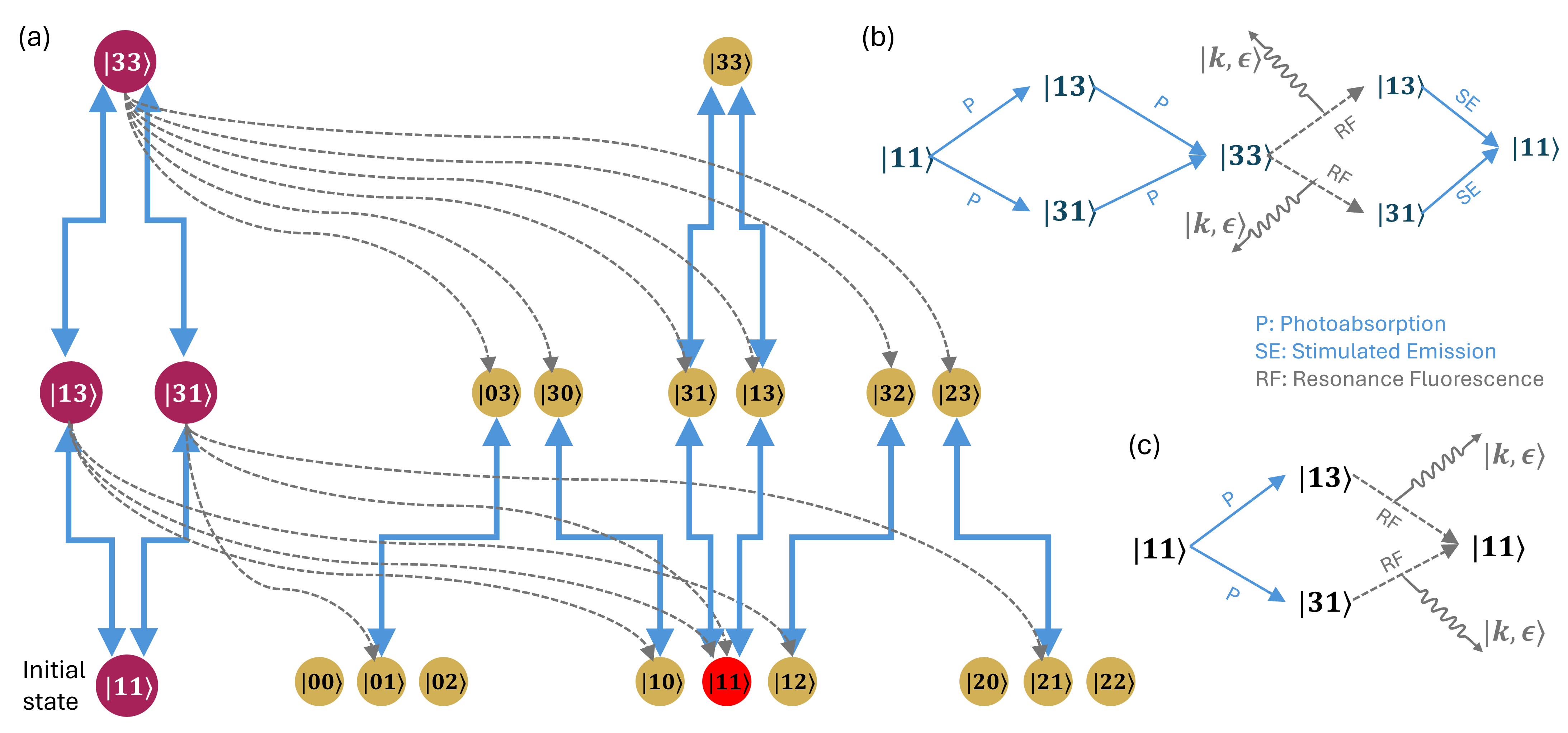}}
\caption{\label{Interference_scheme} Interference pathways for resonance fluorescence from two atoms when the initial state of the system is $\ket{11}$. (a) The states on the left describe the system before emission (unscattered states, purple circles) and the states on the right describe the scattered electronic states (gold circles) of the system entangled with the emitted photon.  The blue solid lines and the gray dashed lines depict Rabi oscillation and resonance fluorescence processes respectively. (b), (c) shows two example pathways to obtain a final scattered state $\ket{11}$ (marked red in (a)) for strong and weak incident intensities, respectively.}
\end{figure*}

We examine the case when the two Ne\textsuperscript{+} ions are both initially in the state with a vacancy in $2p_0$, that is $\ket{\psi_i} = \ket{1s2p_0^{-1}, 1s2p_0^{-1} }$. To explore how scattering yields depend on atomic arrangement, we consider three geometries for the relative orientation vector $\hat{\boldsymbol{R}} = \hat{x}, \hat{y}, \hat{z}$.  In particular, the $\hat{y}$ and $\hat{z}$ cases serve as analogs to Young’s double-slit experiment, with the two Ne\textsuperscript{+} ions acting as scattering centers illuminated by an incident coherent x-ray pulse.

We explore two spatial separations, $R = 10 \lambda_{in}$ and $R\rightarrow0$, where $\lambda_{in} = 2\pi/ k_{in}$ is the incident x-ray wavelength ($\lambda_{in}\sim 27.57$ a.u.). These two limits respectively correspond to cases where the x-ray can and cannot resolve the inter-atomic structure.  Additionally, we examine the system’s response under three different pulse intensities corresponding to pulse areas of $Q = 2\pi$ and $Q = \pi$, representing the two extremes of Rabi oscillations and an intermediate pulse area $Q = 1.5\pi$. A $2\pi$ pulse (intensity $I \sim 2.1\times 10^{18}$ W/cm\textsuperscript{2}) drives a full Rabi cycle—exciting each Ne\textsuperscript{+} ion from $\ket{1}$ to $\ket{3}$ and then back—while a $\pi$ pulse (intensity $I \sim 5.2\times 10^{17}$ W/cm\textsuperscript{2}) inverts the population.

\subsubsection{Q = $2\pi$ pulse}
Figure \ref{Fig_2Pi_signal_initialstate2pz}(a) shows the calculated scattering yields for the $Q = 2\pi$ pulse and atomic separation $\boldsymbol{R} = 10\lambda_{in} \hat{z}$. For simplicity and to restrict the parameter space, we focus on planar scattering where $\boldsymbol{k}$, $\boldsymbol{k}_{in}$ and $\boldsymbol{\epsilon}_{in}$ lie in the same plane(i.e., $\phi_s = 90^{\degree}$, see Fig.~\ref{Schematic_diagram}). The polarization vector $\boldsymbol{\epsilon}$ of the emitted photon need not lie in this plane. The present calculations involve a sum over final photon polarizations [Eq.~(\ref{polarization_choices})].

We first examine the ES yield, shown as orange points in Fig.~\ref{Fig_2Pi_signal_initialstate2pz}(a). ES originates from any occupied state and does not alter the state of the scatterers. The scattering amplitudes from two identical scatterers contain a phase difference of $e^{i\boldsymbol{q}\cdot \boldsymbol{R}}$ [Eq.~(\ref{eqn_psi1_nstate})], giving rise to constructive interference at several $\theta_s$ values that satisfy $\boldsymbol{q}\cdot \boldsymbol{R}= 2 \alpha \pi$ for $\alpha \in $ Integers.  This interference arises because an elastically scattered photon emerging from the two atoms in state $\ket{\psi_m, \psi_n}$, cannot be traced to a specific atom here. We compare the calculated ES yield to a fit that is proportional to the structure factor for Thomson scattering from two atoms,
\begin{equation} \label{fit_function}
    g(\theta_s) \propto \frac{d \sigma_{th}}{d \Omega} ~ \abs{f_1(\boldsymbol{q})}^2 ~ \abs{ 1 + e^{i\boldsymbol{q}\cdot \boldsymbol{R}} }^2,
\end{equation}
where $\abs{f_1(\boldsymbol{q})}^2$ is the ES atomic form factor of Ne\textsuperscript{+} for the state $1s2p_0^{-1}$ (Sec.~\ref{Methods}). For the given incident photon energy, $f_1(\boldsymbol{q}) \approx f_3(\boldsymbol{q})$ for all angles. The computed elastic yield matches this fit (Fig.~\ref{Fig_2Pi_signal_initialstate2pz}(a)), indicating that the ES channel captures the atomic structure consistent with expectations from non-resonant x-ray diffraction. 

Next, we examine the RF channel, sometimes referred to as ``resonance scattering"~\cite{Sakurai_adv}. In comparison to ES pathway, the RF pathway has a higher yield and shows similar angular dependence but with a smaller fringe contrast, as shown in Fig.~\ref{Fig_2Pi_signal_initialstate2pz}(a).  For a single atom, the angular dependence of RF is isotropic when the outgoing photon polarization is not measured~\cite{Res_theory_PRA}. However the interference in a small fraction of RF pathways from the two-atom system introduces a slight anisotropy. Due to polarization imposed selection rules, the bound-state population during the pulse is effectively restricted to the $1s2p_0^{-1}$ and $1s^{-1}2p$ states. A single Ne\textsuperscript{+} ion emits an RF photon only if it has a nonzero population in its core-hole state $\ket{3}$ (Fig.~\ref{Fig_ElectronicStates}). Thus, the two-atom system emits an RF photon when occupying states \{$\ket{13}$,$\ket{31}$,$\ket{33}$\}.  Depending on $\boldsymbol{k}$ and $\boldsymbol{\epsilon}$ of the RF photon and if further photons are absorbed, the two-atom system may evolve into any of 16 possible final states. These RF pathways are illustrated schematically in Fig.~\ref{Interference_scheme}. At the beginning of the pulse, the two atom system is in the initial state $\ket{11}$ shown on the bottom-left in Fig.~\ref{Interference_scheme} (a). During the pulse, in the absence of RF, one or both atoms can get excited or de-excited during Rabi oscillation shown as blue arrows and states denoted using purple circles. If an RF photon is emitted, the emitting atom transitions to one of the degenerate ground states ${\ket{0}, \ket{1}, \ket{2}}$ (depicted by gray dashed lines), yielding a set of final states (gold circles) entangled with the outgoing photon. Additional Rabi cycling of these states may follow. Figures.~\ref{Interference_scheme}(b) and ~\ref{Interference_scheme}(c) depict two possible RF interference pathways for the strong- and weak-field cases, respectively.

Fig.~\ref{Fig_2Pi_signal_initialstate2pz}(b) shows the angular distribution of RF contributions from different final scattered states. Only $\ket{11}$ final scattered state exhibits significant interference fringes resembling those of ES. Figures~\ref{Fig_2Pi_signal_initialstate2pz}(c) and \ref{Fig_2Pi_signal_initialstate2pz}(d) reveal the contribution from each final scattered state for two scattering angles of 3$\degree$ and 26.75$\degree$, corresponding to the interference minima in panel (b) of Fig.~\ref{Fig_2Pi_signal_initialstate2pz}. These figures confirm that the final scattered state $\ket{11}$ exhibits the strongest position dependence, indicative of significant quantum interference. The $\ket{13}$ and $\ket{31}$ scattered states show a small position dependence as well. Importantly, the limit $\boldsymbol{R} = 0$ corresponds to the case of maximum constructive interference at any given scattering angle. In contrast, final scattered states involving one atom in $\ket{0}$ or $\ket{2}$ exhibit no dependence on the interatomic separation $\boldsymbol{R}$ and thus show no interference. This occurs because these states can only be reached through RF if that atom had been excited to the core-hole state $\ket{3}$. Therefore, when an atom ends in $\ket{0}$ or $\ket{2}$, the origin of the photon becomes distinguishable, precluding interference.  Among the final states, $\ket{11}$, $\ket{13}$, $\ket{31}$, and $\ket{33}$ can each be reached via multiple indistinguishable RF pathways. As a result, they exhibit varying degrees of interference depending on the transition amplitudes of the RF pathways.

Notably, Figs.~\ref{Fig_2Pi_signal_initialstate2pz}(b)–(d) reveal that the $\ket{11}$ final state can exhibit perfect destructive interference in the RF channel. This occurs only when the emission amplitudes from the two atoms are equal in magnitude and opposite in phase, leading to complete cancellation. Physically, this implies that it is equally probable for either atom to have emitted the RF photon. A conceptual explanation for this behavior lies in the nature of the $Q = 2\pi$ pulse: under such conditions, the system would return to the $\ket{11}$ state even in the absence of fluorescence emission. Hence, if the system ends in $\ket{11}$ after emitting an RF photon, there is intrinsic ambiguity in identifying which atom emitted it, which is an essential requirement for quantum interference.

We note that the final scattered states with at least one atom in $\ket{3}$ predominantly undergo Auger decay ($\sim\!\!99\%$ probability) after the pulse ends \textemdash a process not explicitly modeled in this work.  This is justified for two reasons. First, due to the low scattering cross sections, the probability for an atom to scatter a photon and end up after the pulse in a core-hole state is about 2\% or less, making its impact on the population dynamics negligible. Second, after the system has scattered a photon, any subsequent decay process in the scattered states does not change the photon yield.

We now turn to the the total response from both ES and RF channels.  Fig.~\ref{Fig_2Pi_signal_initialstate2pz}(a) shows the angular distribution of the photon yield obtained by retaining all source terms from both channels in Eq.(\ref{eqn_psi1_nstate}), labeled as the ``Coh sum". For comparison, the corresponding incoherent sum is also shown which is defined as the sum of individual channel probabilities, in contrast to the square of the summed amplitudes used in the coherent case. The agreement between the incoherent sum and coherent sum indicates the absence of interference effects in the angular distribution between the two channels for the chosen initial state. This result is consistent with our prior findings for single-atom responses~\cite{Res_Rabi_PRAL, Res_theory_PRA}. Interestingly, the angular dependence of the coherent sum closely resembles that of the ES signal, though it is shifted and slightly enhanced. This observation is supported by the agreement with the fitted structure factor function $g(\theta_s)$ [see Eq.~(\ref{fit_function})], shown as a solid black line in Fig.~\ref{Fig_2Pi_signal_initialstate2pz}(a), with an appropriate vertical shift applied. 

Additional results for three alternative atomic arrangements are presented in Fig.~\ref{Fig_2Pi_signal_initialstate2pz}(e). For $\boldsymbol{R}=10\lambda_{in}\hat{x}$, the different periodicity of the fringes i.e. the distinct scattering angle dependence is a consequence of the different structure factor associated with this arrangement. The case of $\boldsymbol{R}=10\lambda_{in}\hat{y}$ does not have any fringes because $\boldsymbol{q} \cdot \boldsymbol{R} = 0$ for planar scattering . 
At small scattering angles, the angular dependence of the coherent sum yield can in principle be used to extract the inter-atomic structure. At large scattering angles, while the fringes remain, the amplitudes of the fringes decreases. Note that the RF channel consists of both interfering pathways and non-interfering pathways. As the scattering angle increases, $\epsilon \cdot \epsilon_{in}$ decreases and hence the contribution from distinguishable final scattered states increases. Therefore the non-interfering part of the RF increases. This non-interfering part of RF has sometimes been referred to as ``incoherent component" in some previous works~\cite{itano1998_Youngs, Richter1991_RFinterference}.

\begin{figure*}
\resizebox{180mm}{!}{\includegraphics {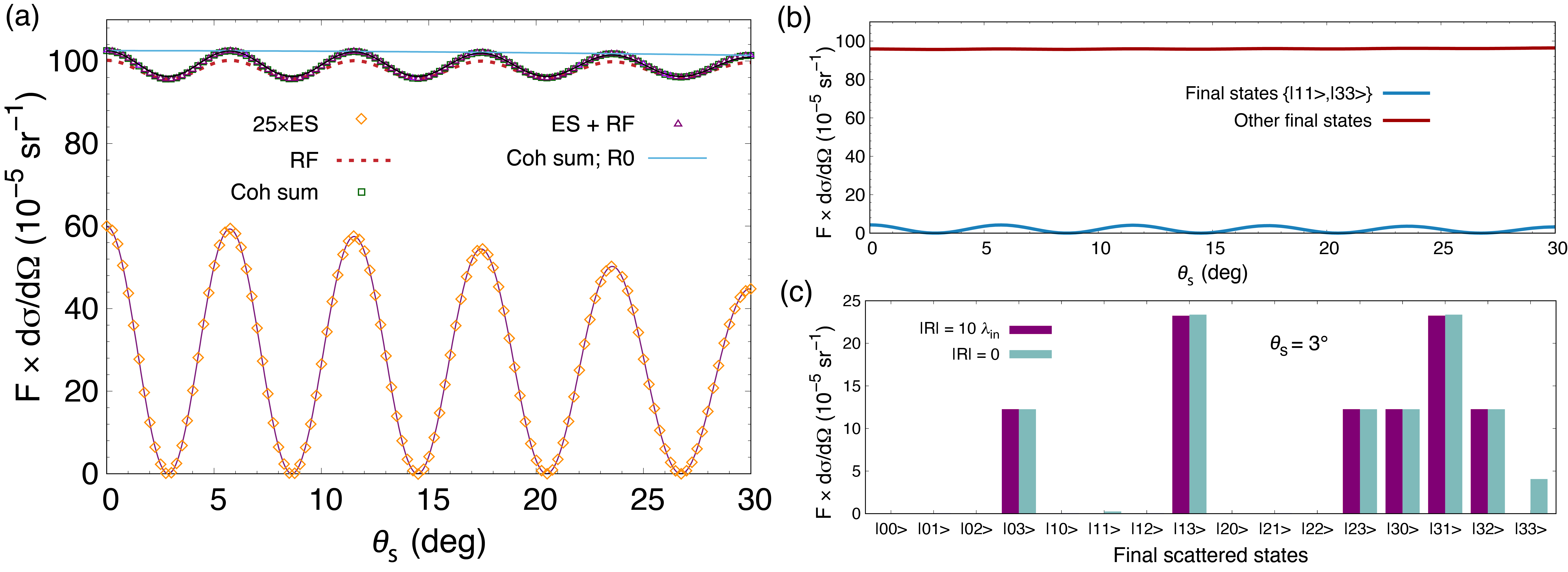}}
\caption{ Resonant scattering photon yield from two atoms using a 0.25 fs ($t_{wid}$) pulse with $Q = \pi$. (a) Scattering angle dependence from the different channels. Note that ES yield is multiplied by a factor of 25 to show its angular dependence. (b) Scattering angle dependence of resonance fluorescence for different final scattered states. (c) Position dependence of resonance fluorescence yields from different final scattered states for $\theta_s = 3\degree$. The other parameters are the same as Fig.~\ref{Fig_2Pi_signal_initialstate2pz}.
}
\label{Fig_Pi_signal_initialstate2pz}
\end{figure*}

\subsubsection{Q = $\pi$ pulse}
Next, we examine the scattering response in a $\pi$ pulse, as shown in Fig.~\ref{Fig_Pi_signal_initialstate2pz}. In the absence of RF, such a pulse would transfer the population from the ground state $\ket{1}$ to the core-hole state $\ket{3}$ at the end of pulse, with some probability for ionization during the pulse. Therefore,  as a result, the system produces significant RF after the pulse. The ES channel even though is present during the pulse, its yield is about two orders of magnitude smaller compared to the RF channel. Nonetheless, its angular dependence closely resembles that observed for the $Q = 2\pi$ case. This is supported by agreement using the same fit function of Eq.~(\ref{fit_function}) but with a different proportionality constant. The comparison of 
 Fig.~\ref{Fig_Pi_signal_initialstate2pz} with Fig.~\ref{Fig_2Pi_signal_initialstate2pz} shows that the RF signal for $Q=\pi$ displays reduced angular modulation (fringe contrast) relative to the $Q=2\pi$ case.

As in the $Q=2\pi$ case, the coherent sum of the ES and RF channel agrees with their incoherent sum as expected. For small-angle scattering, the coherent sum is also found to exhibit qualitatively similar scattering angle dependence to that of ES contribution as evidenced by agreement with a y-shifted fit function.

Figure ~\ref{Fig_Pi_signal_initialstate2pz}(b) explains why the RF contribution in Fig.~\ref{Fig_Pi_signal_initialstate2pz}(a) only exhibits a small scattering angle dependence. For $Q=\pi$ both atoms get excited and therefore the final scattered states that manifest are those where one of the two excited atoms emits a photon. The final scattered states $\ket{03}, \ket{30}$, $\ket{23}$, $\ket{32}$ have no interference as they uniquely identify the emitter. Hence these have no position dependence and this is evident in Fig.~\ref{Fig_Pi_signal_initialstate2pz}(c). Additionally, one can understand the maximum interference in the case of $\ket{33}$ using the previous conceptual argument. That is, for the given pulse with $Q = \pi$, each atom (bound part) would have ended up in the final state $\ket{3}$ at the end of the pulse, even if it had not emitted an RF photon. Therefore for the given final scattered state $\ket{33}$, either atom is equally probable to have emitted the RF photon.

Our results suggest that, for a given pulse duration, the intensity of the pulse can be used to control fringe contrasts. It is worth pointing out the dependence of the fringe contrast on the pulse area is analogous to the time-dependence of the first-order interference effects in RF reported for monochromatic fields by Richter~\cite{Richter1991_RFinterference}. For imaging applications, given the attosecond time-scale, it can be challenging to develop a time-resolving photon detector. On the other hand, controlling fringe contrast through controlling pulse area by tuning the incident intensity may be a better useful alternative. In addition, the fringe contrasts can also be improved by selecting the polarization of outgoing photons which helps narrow the range of final scattered states to those which allow for interference in RF between two atoms. For example, the final scattered states exhibiting interfering RF pathways are  $\ket{11}$, $\ket{13}$, $\ket{31}$, and $\ket{33}$. For planar scattering ($\phi_s = 90\degree$), these arise from outgoing photons with polarization in the scattering plane [Eq.~(\ref{polarization_choices})] and when $\epsilon \cdot \epsilon_{in} \neq 0$. The idea of selecting outgoing photon polarization to improve fringe contrast has been experimentally demonstrated for weak-field resonant scattering in optical regime~\cite{itano1998_Youngs, Eichmann_Youngs_1993}.

\subsubsection{Q = 1.5 $\pi$ pulse}
\begin{figure*}
\resizebox{180mm}{!}{\includegraphics {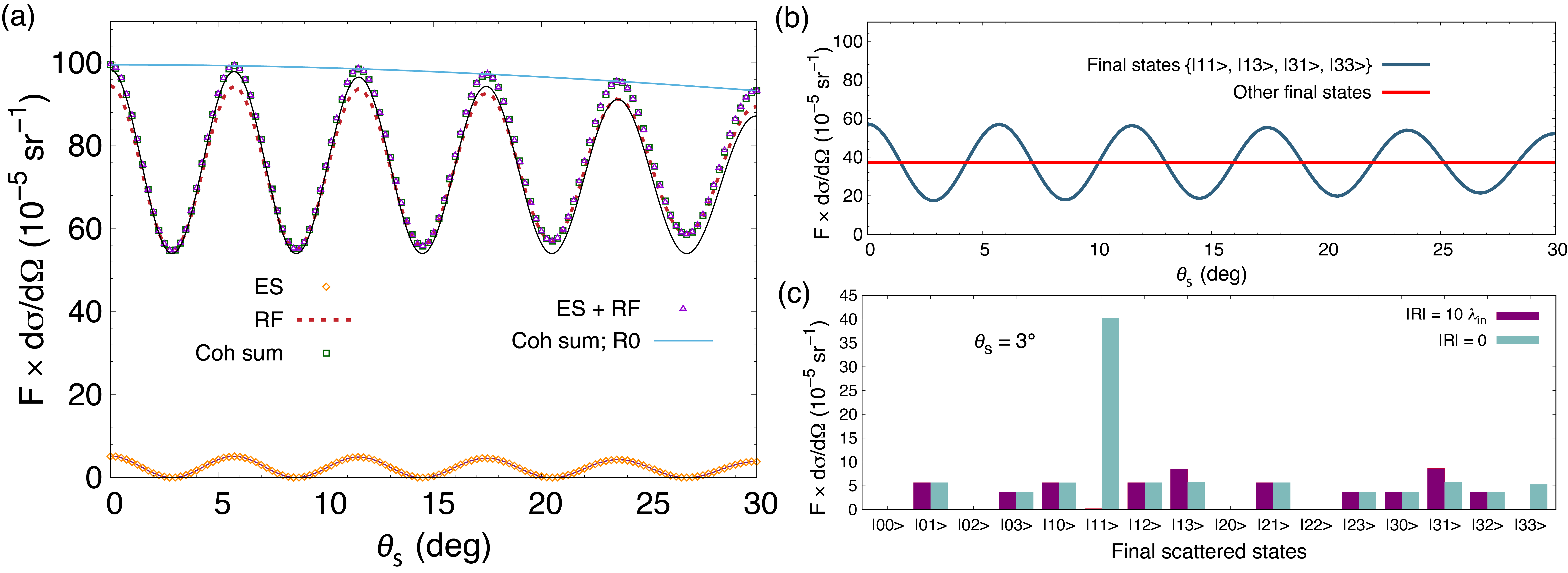}}
\caption{ Resonant scattering photon yield from two atoms using a 0.25 fs ($t_{wid}$) pulse with $Q = 1.5\pi$. (a) Scattering angle dependence from the different channels. (b) Scattering angle dependence of resonance fluorescence for different final scattered states. (c) Position dependence of resonance fluorescence yields from different final scattered states for $\theta_s = 3\degree$. The other parameters are the same as Fig.~\ref{Fig_2Pi_signal_initialstate2pz}.
}
\label{Fig_1p5Pi_signal_initialstate2pz}
\end{figure*}

We examine the resonant response for a pulse area of $Q = 1.5\pi$. This intermediate pulse drives 75\% of a Rabi cycle and, in the absence of photoionization and inner-shell decay channels, would leave the system in an equal coherent superposition of the ground and core-excited states. 

The calculated scattering yields are shown in Fig. \ref{Fig_1p5Pi_signal_initialstate2pz}. The ES scattering yield, which depends approximately linearly on intensity, is larger than the ES in $Q = \pi$ but smaller than that of $Q =2\pi$ (which corresponds to a pulse intensity four times larger than that of the $\pi$-pulse case). The RF yield, however, is substantially larger than that of $Q = 2\pi$ but remains below that of $Q = \pi$. This behavior arises because a significant population remains in the excited state at the end of the pulse, leading to a substantial amount of post-pulse RF emission.  Such emission is absent in the $Q=2\pi$ case, 
where the system is fully driven back to the ground state, but remains smaller than in the $Q=\pi$ case due to the reduced excited-state population. The interference in the RF channel can be understood from earlier indistinguishability arguments. In the absence of RF, each atom would end up in either $\ket{1}$ or $\ket{3}$. When an RF photon is emitted, the resulting final scatttered states $\ket{11}$, $\ket{13}$, $\ket{31}$, and $\ket{33}$ exhibit varying degrees of interference with maximum interference exhibited by $\ket{11}$ and $\ket{33}$ as for each of these states, it is equally probable for either of the atoms to have emitted the photon.

When both ES and RF channels are included, the total yield for this case is found to exhibit a fringe contrast which is significantly higher than $Q = \pi$, but lower than for $Q = 2\pi$. Intermediate pulse areas are therefore expected to follow similar trends in the individual channel yields and in the total yields. In Appendix~\ref{App_RFtrendintensity}, we provide additional calculations showing 
that the interference in the RF channel decreases 
systematically with increasing intensity, consistent with previous observations 
for monochromatic optical fields~\cite{Richter1991_RFinterference,agarwal2002_Iorder_vs_IIorder,Keitel2007_strongfield_interf}.

\subsubsection{Contribution of Multiple Resonances and Ne$^{2+}$ to scattering probability} \label{Subsec_Ion_contrib}

\begin{figure}
%\resizebox{70mm}{!}{\includegraphics{I_rev/Signal_2atoms_2pzonly_Ioncontrb.eps}}
%\resizebox{70mm}{!}{\includegraphics{I_rev/1p5_Pi_Signal_2atoms_2pzonly_Ioncontrb.eps}}
\resizebox{85mm}{!}{\includegraphics{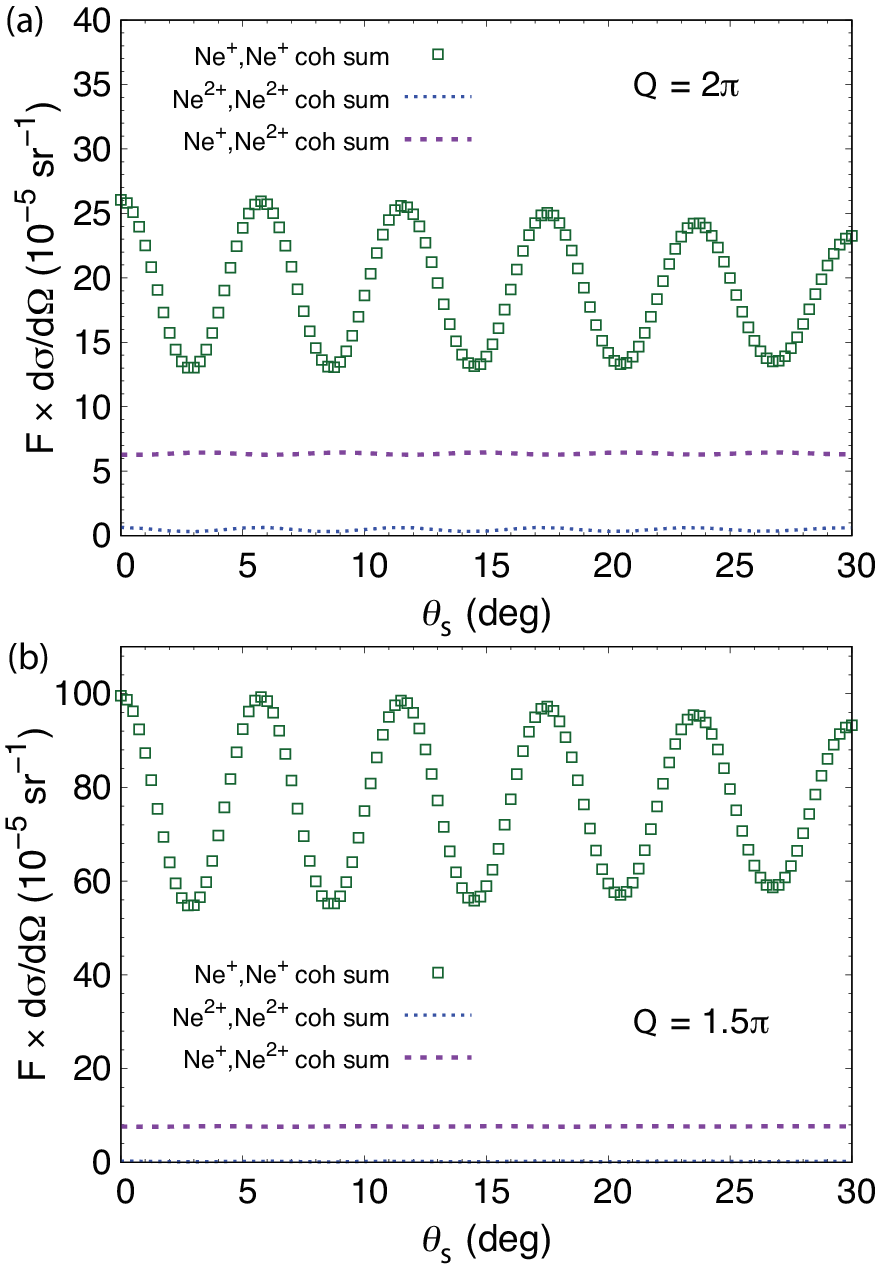}}
\caption{\label{Fig_Ioncontribution}
Estimate for scattering contribution from higher charged ions created during the pulse for the cases $Q=2\pi$ and $Q = 1.5\pi$, respectively.
}
\end{figure}

During resonant driving of Ne\textsuperscript{+}, further ionization produce 
higher-charge states. In general, these species need not be resonant with the incident pulse. In this case, the 1s-2p transition of Ne\textsuperscript{2+} with $2p^{-2}$ and $2s^{-1}2p^{-1}$ are detuned by approximately $\sim$ 7.7 and 6.5 eV respectively from the incident photon energy. Further, the Ne\textsuperscript{2+} can be produced in a range of initial states, $1s^{-2}$, $1s^{-1}2s^{-1}$, $1s^{-1}2p^{-1}_{0}$, $1s^{-1}2p^{-1}_{\pm1}$, $2s^{-2}$, $2s^{-1}2p^{-1}_{0}$, $2s^{-1}2p^{-1}_{\pm1}$, $2p^{-2}_{0}$, $2p^{-1}_{0}2p^{-1}_{\pm1}$, $2p^{-2}_{\pm1}$, and 
$2p^{-1}_{1}2p^{-1}_{-1}$. A detailed analysis of their population dynamics is provided in Appendix~\ref{App_dynamics_higherchargedstates}.

Modeling the scattering response of Ne\textsuperscript{2+} may nevertheless require a resonant treatment. This is because the present pulse has a duration of 
0.25~fs, corresponding to a bandwidth of $\sim 7.3~\mathrm{eV}$ (FWHM). Second, even for an ideal monochromatic incident field (zero spectral bandwidth), the strong field can overcome the effects of detuning when the Rabi frequency is larger than detuning.  As a result, the produced Ne\textsuperscript{2+} may exhibit quasi-resonant excitation depending on the initial state and emit a photon through resonance fluorescence. This significantly complicates the dynamics. However, the fluorescence photons from Ne$^{2+}$ appear at energies different from the incident photon energy, so for Ne$^{+}$-Ne$^{2+}$ pairs the RF channel contributes primarily to an angularly uniform background with little interference in the photon yield.

To estimate the contribution from higher-charge species to the total scattering 
signal, we perform two sets of scattering response calculations (Sec.~\ref{Methods}) under identical pulse conditions, initially assuming the system consists of Ne\textsuperscript{+}Ne\textsuperscript{2+} ($\ket{2p^{-1}_{0},2p^{-2}_{0}}$) and Ne\textsuperscript{2+}Ne\textsuperscript{2+} ($\ket{2p^{-1}_{0},2p^{-2}_{0}}$) as the core-excitation from this Ne\textsuperscript{2+} state has the largest transition dipole moment. The resulting photon yields are normalized by the probability at the peak of the pulse for one of the two  Ne\textsuperscript{+} to become ionized to Ne\textsuperscript{2+} and for both Ne\textsuperscript{+} to become ionized to Ne\textsuperscript{2+}, respectively. Only a subset of the Ne\textsuperscript{2+} states produced can contribute to resonance fluorescence, and they have different transition moments and slightly different transition energies. The other Ne\textsuperscript{2+} states which do not contribute to RF can scatter a photon through Thomson scattering. Using the total Ne$^{2+}$ population therefore provides an upper bound on their possible contribution from during the pulse. Because Ne$^{2+}$ can be created at any instant during the pulse, each ion experiences a different effective intensity, making their collective interaction with the field effectively incoherent.
%However using the entire population of Ne\textsuperscript{2+} for the estimate for the scattering contribution provides a useful upper-bound. It should be noted that this is only an estimate as these species have a non-zero probability to be created at any instance during the pulse making the intensity experienced by Ne\textsuperscript{2+} depend on when it is created, thus making their interaction with the pulse effectively incoherent as each Ne\textsuperscript{2+} created experiences a different excitation probability. The results in this section serves as an estimate for the upper-bound for the resonant scattering contribution from Ne\textsuperscript{2+} created during the pulse.

Fig.~\ref{Fig_Ioncontribution} summarizes the results for the two pulse areas $Q = 2\pi$ and $Q = 1.5\pi$. Three conclusions emerge. First, the total scattered yield from Ne$^{+}$-Ne$^{2+}$ pairs shows very little interference. Although ES pathways from the two species exhibits interference, the dominant RF contribution shows almost no interference, resulting in an angularly nearly constant background. 
Second, the contribution from Ne$^{2+}$-Ne$^{2+}$ pairs is negligible because the probability of forming two Ne$^{2+}$ ions \emph{and} having both scatter a photon is very small. Third, when such scattering does occur, the Ne$^{2+}$--Ne$^{2+}$ channel exhibits noticeable interference with fringe patterns qualitatively similar to those of Ne$^{+}$-Ne$^{+}$, although its overall magnitude is far smaller. 

These results suggest that in samples where several ionic species have nearby resonances, this may add to substantial background in experiments. If the resonances are very close to the incident photon energy, for intense pulses, the scattering pattern may become difficult to interpret.

\begin{figure*}[t]
\resizebox{180mm}{!}{\includegraphics {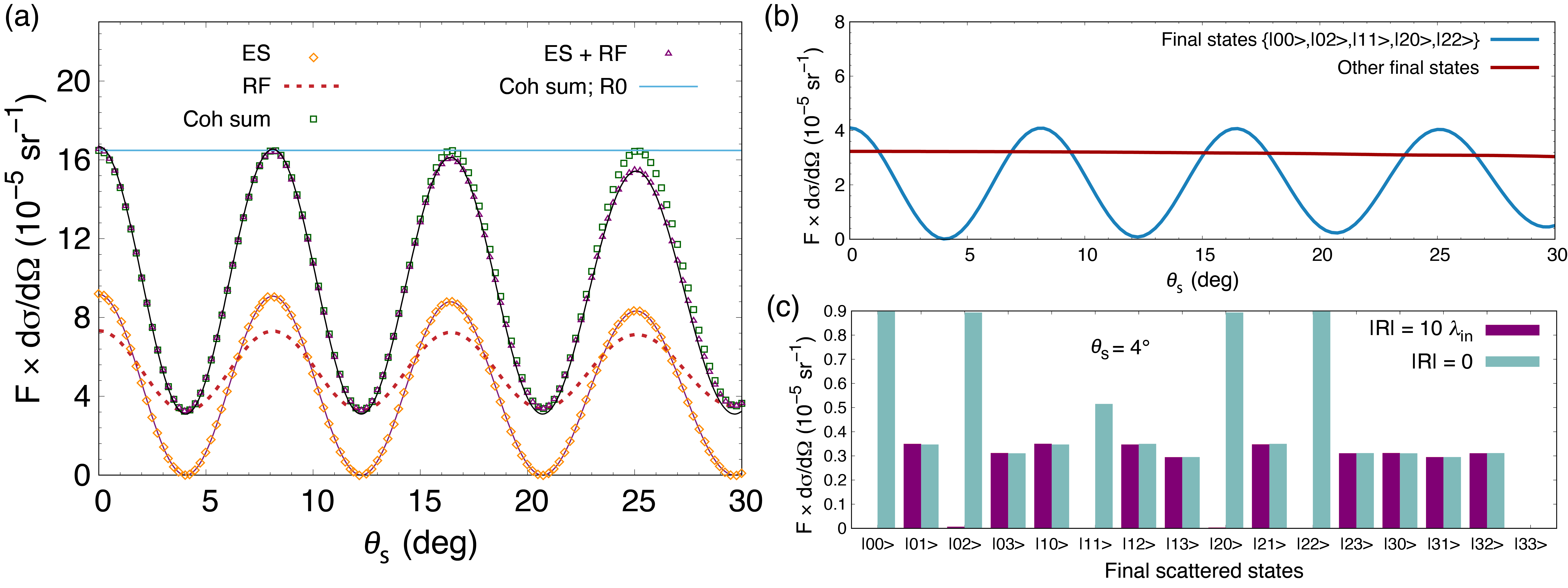}}
\caption{Resonant scattering photon yield from two atoms when both are initially in the same superposition state $\ket{\psi_{sup}} = \frac{1}{\sqrt{3}} \big(\ket{0} + \ket{1} + \ket{2}\big)$ [Eq.~(\ref{Eqn_initialstate_superposition})] and azimuthal angle $\phi_s =135 \degree$ (out of the plane). (a) Scattering angle dependence for different scattering channels. (b) Resonance fluorescence dependence on scattering angle and final scattered state. (c) Position dependence of resonance fluorescence yields from different final scattered states for $\theta_s = 4 \degree$. The other parameters are the same as Fig.~\ref{Fig_2Pi_signal_initialstate2pz}.
}
\label{Fig_2Pi_signal_sup_equal_phi135}
\end{figure*}

\subsection{An equal superposition initial state} \label{subsec_initial_supequal}
% Figure Q=2pi initial state equal superposition

We investigate the case when the initial state of each Ne\textsuperscript{+} ion is an equal superposition of the three degenerate states $\ket{\psi_{sup}} =  \frac{1}{\sqrt{3}} \big[ \ket{ 1s2p_{-1}^{-1} } + \ket{ 1s2p_0^{-1}} + \ket{ 1s2p_{+1}^{-1} } \big]$ for an incident pulse with $Q = 2\pi$. The initial state of the system is given by,
\begin{equation} \label{Eqn_initialstate_superposition}
    \ket{\psi_i} = \ket{\psi_{sup}, \psi_{sup}}
\end{equation}
In our previous work, for the same pulse condition, the single-atom response of an equal superposition initial state was found to exhibit substantial interference between the ES and RF channels~\cite{Res_theory_PRA} at large scattering angles ($\theta_s$), particularly for azimuthal angle (see Fig.~\ref{Schematic_diagram}) $\phi_s=135\degree$ (constructive interference) and $\phi_s=45\degree$(destructive interference).

Figure~\ref{Fig_2Pi_signal_sup_equal_phi135} shows the two-atom resonant scattering response for this initial state [Eq.~(\ref{Eqn_initialstate_superposition})] for $\phi_s = 135\degree$ which is $45\degree$ off the plane containing $\boldsymbol{k}_{in}$ and $\boldsymbol{\epsilon}_{in}$.
The ES channel displays a similar angular dependence as before, with the interference minimum slightly shifted by $\approx 1 \degree$ due to changes in the momentum transfer vector $\boldsymbol{q}$. The overall RF yield is reduced because only one-third of the population in each atom participates in the Rabi dynamics, limiting the maximum core-hole population accordingly. However, the RF angular modulation appears more pronounced than for the initial state $\ket{ 1s2p_0^{-1}}$ [compare Fig.~\ref{Fig_2Pi_signal_sup_equal_phi135}(a) with Fig.~\ref{Fig_2Pi_signal_initialstate2pz}(a))]. The contribution to the RF yield from different final scattered states, as shown in Fig.~\ref{Fig_2Pi_signal_sup_equal_phi135}(b) and Fig.~\ref{Fig_2Pi_signal_sup_equal_phi135}(c) for $\theta_s = 4\degree$ (destructive interference), reveals substantial interference in the RF pathways between the two atoms for several final scattered states. This is expected for these final scattered states, as the initial superposition state [Eq.(\ref{Eqn_initialstate_superposition})] prevents one from attributing the emitted RF photon to a specific atom.

Notably, the coherent sum deviates from the incoherent sum at large scattering angles, indicating interference between the ES channel and RF channels within a single atom, consistent with our previous results~\cite{Res_Rabi_PRAL, Res_theory_PRA}. The periodicity (minimum and maximum) of the coherent sum resembles the periodicity in ES, suggesting that the coherent sum still reflects the structure. However, the exact shape of the coherent sum starts to deviate from that of ES yield for large scattering angles.

\subsection{Linear scattering regime} \label{subsec_weakfield}
\begin{figure}[!h]
\resizebox{85mm}{!}{\includegraphics {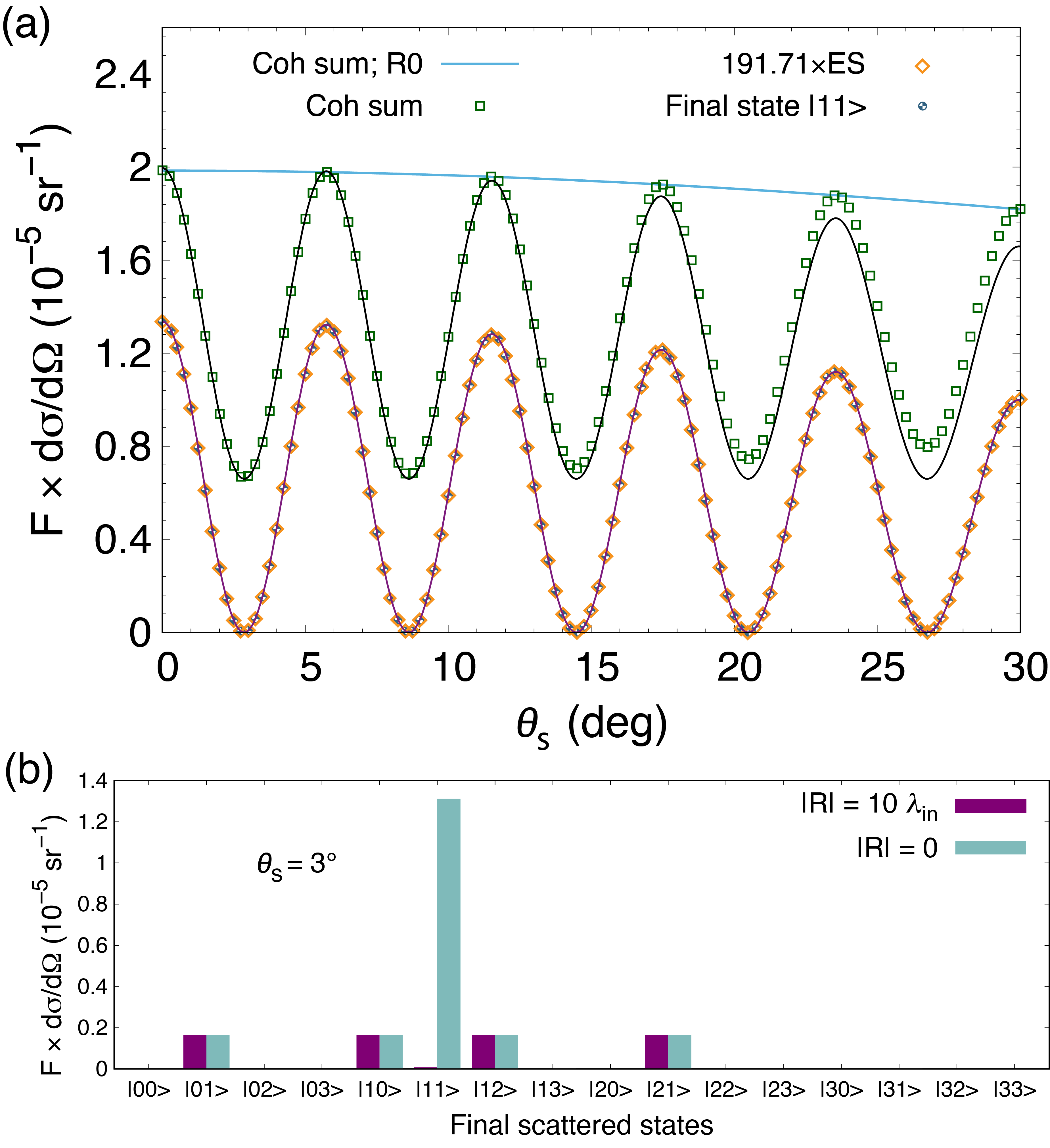}}
\caption{Resonant scattering photon yield from two atoms for a low intensity incident pulse (Q$\sim$0.16) of duration 0.25 fs ($t_{wid}$). The initial state of each atom is $\ket{1}$. (a) Scattering angle dependence from different channels. The contribution to the photon yield from only resonance fluorescence into the final scattered state $\ket{11}$ is denoted as `Final state $\ket{11}$'. Note that ES yield is multiplied by a factor of 191.71 to show its angular dependence. (b) Position dependence of resonance fluorescence yields from different final scattered states. The other parameters are the same as Fig.~\ref{Fig_2Pi_signal_initialstate2pz}.
}
\label{Fig_Ep2_signal_initialstate_2pz}
\end{figure}

We now examine the resonant x-ray scattering response from the two-atom system exposed to a $0.25$-fs pulse of an intensity which corresponds to the linear scattering regime. In this regime, the incident pulse is not intense enough to transfer any significant population from the ground state $1s2p_0^{-1}$ to the core-hole excited state $1s^{-1}2p$ and therefore causes no Rabi cycling. However, the incident pulse can still scatter (both ES and RF) from the system or photoionize it. For this case, the initial state of the system is chosen to be $\ket{1s2p_0^{-1}, 1s2p_0^{-1}}$.

Figure~\ref{Fig_Ep2_signal_initialstate_2pz}, shows the resulting scattering response for a pulse intensity of about $1.4\times10^{15}~$W/cm\textsuperscript{2}, which corresponds to a pulse area of $Q\sim0.16$. At this intensity, the ES contribution is two orders of magnitude smaller than the RF channel. Therefore, the coherent sum is effectively the same as the RF yield. While the coherent sum exhibits the same angular periodicity as ES, an exact match with ES is only observed at small scattering angles. This deviation is reflected in the mismatch with the shifted fit function $g(\theta_s)$ of Eq.~(\ref{fit_function}).

In this intensity regime, the resonant scattering process can be interpreted as a one-photon in one-photon out process (see Fig.~\ref{Interference_scheme}(c)). That is, only one of the two atoms is likely to be excited during the pulse, and after emission of an RF photon,  the system returns to a final state that does not contain any core-excited population. Depending on the polarization of the outgoing photon, several final scattered states $\ket{01}$, $\ket{10}$, $\ket{11}$, $\ket{12}$, and $\ket{21}$ are accessible. However only the final scattered state $\ket{11}$ exhibits interference, as in this case one cannot determine which of the two atoms scattered the photon [Fig.~\ref{Fig_Ep2_signal_initialstate_2pz}(b)]. The resulting interference pattern from this RF pathway is qualitatively identical to that of the ES channel, as evident in Fig.\ref{Fig_Ep2_signal_initialstate_2pz}(a).

These results suggests that in the linear ultrafast regime, resonant x-ray scattering can reveal the same structural information as conventional non-resonant x-ray scattering, but with a substantially increased yield relative to non-resonant scattering at the same intensity. In addition in this intensity regime, the ionization through photoionization and Auger-decay pathways is found to be negligible and hence the sample damage is minimal.

\section{Conclusion and summary} \label{conclusion_summary}
In this work, we described a theoretical approach for investigating the resonant x-ray scattering response from two identical non-interacting atoms exposed to an intense attosecond pulse for different intensities, positions of the two atoms, and their initial state. Our calculations showed that while elastic scattering pathways from the two atoms interfere, only a fraction of the resonance fluorescence pathways from the two scatterers can lead to interference. This interference is sensitive to the relative atomic positions and therefore encodes structural information.

Remarkably, we found that the total scattered yield qualitatively reproduces the structure factor known from non-resonant x-ray diffraction. This indicates that resonant x-ray scattering with attosecond pulses, despite being driven by fundamentally different processes,  can reveal equivalent interatomic structural features. For the intensities presented, resonant scattering produces a total photon yield that exceeds its non-resonant counterpart by at least a factor of two or more.  Both the fringe visibility and the degree of yield enhancement depends strongly on the pulse area and the initial electronic state of the system.
A $\pi$ pulse produces the highest photon yield, but has the lowest fringe visibility.  On the other hand, the highest fringe visibility were obtained in linear scattering regime, where ionization is minimal and no Rabi cycling occurs. This regime not only maximizes the ratio of resonant to non-resonant yield for that intensity but also provides optimal structural sensitivity. This finding is consistent with previous predictions made in the context of monochromatic optical fields~\cite{Richter1991_RFinterference, Keitel2007_strongfield_interf}.  However, the photon yield for $Q\sim0.16$ is about 1 factor of 50 lower than that of a $\pi$ pulse.

For moderately strong fields associated with few Rabi oscillations, the fringe contrast and the photon yield can be improved by controlling the pulse area (tuning intensity for a given pulse duration). Additional control over the polarization of the outgoing photons can further enhance fringe contrast. Notably, if the two atoms are initially in an superposition state, interference is improved due to the presence of more indistinguishable RF pathways.

Our results also offer insights into how the ultrafast resonant scattering response scales with system size. In the forward direction, ES signal is expected to scale approximately with the square of the number of atoms, provided ionization remains low, mirroring the scaling behavior familiar from conventional x-ray diffraction. For reference, inelastic processes (such as Compton scattering) are typically incoherent and scale linearly with the number of scatterers. Interestingly, our calculations suggest that under conditions where interference is prominent, the total resonant scattering yield in the forward direction can exceed linear scaling and approach a quadratic dependence in the linear scattering regime.

Overall, these findings highlight the potential of ultrafast resonant x-ray scattering with intense pulses as a structural probe that combines high photon yield with enhanced elemental sensitivity. Since terawatt-scale hard x-ray pulses are now feasible at facilities such as the European XFEL~\cite{yan2024terawatt}, future studies of resonant scattering studies in heavy-element systems using intense attosecond hard x-ray pulse may be insightful. In this regime, inner-shell decay channels are suppressed, valence ionization is less likely, and impact ionization from photoelectrons becomes inefficient due to their high kinetic energy. Together, these effects may enable nearly damage-free, element-specific imaging at atomic resolution.

\section{Acknowledgements}
A.V. thanks E. Pelimanni for suggestions about visualization. This work was supported by the U.S. Department of Energy, Office of Basic Energy Sciences, Division of Chemical Sciences, Geosciences, and Biosciences through Argonne National Laboratory. Argonne is a U.S. Department of Energy laboratory managed by UChicago Argonne, LLC, under Contract No. DE-AC02-06CH11357. We gratefully acknowledge the computing resources provided on Improv, a high-performance computing cluster operated by the Laboratory Computing Resource Center at Argonne National Laboratory.

\section{Data Availability}
The data plotted in the figures and the atomic data used in the 18-level density matrix calculation will be openly available upon publication at Ref.~\cite{Figure_data}. %Zenodo ( 10.5281/zenodo.15604377).
%The data plotted in the figures are available in Ref.?
%\appendix
%\section{}\label{App_}

\appendix

\section{Coherent Dynamics of Higher Charge States} \label{App_dynamics_higherchargedstates}

\begin{figure*}
\resizebox{175mm}{!}{\includegraphics{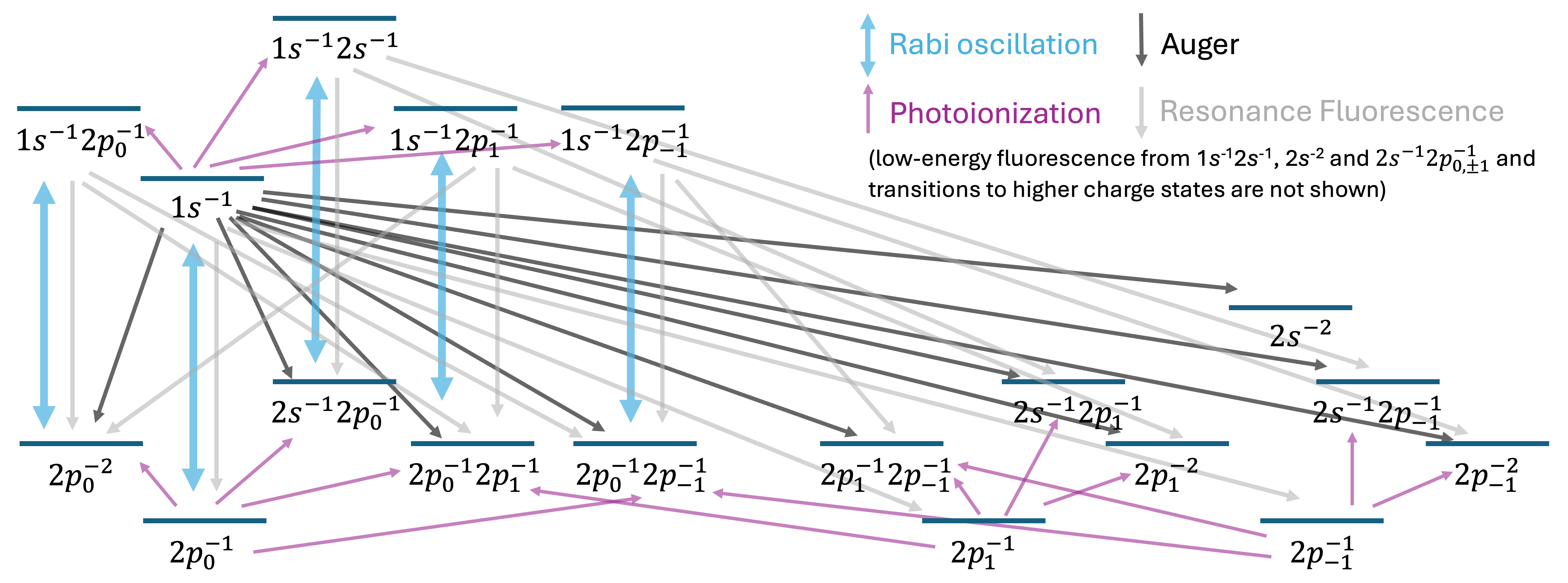}}
\caption{\label{Fig_18levels} Electronic transitions of 18 electronic configurations of Ne\textsuperscript{+} and Ne\textsuperscript{2+} used in the master equation for describing the coherent and relaxation dynamics involving five resonances. The figure includes photoionization (purple arrow) and Auger decay (dark gray arrows) pathways and electronic transitions  associated with Rabi oscillations (blue arrows) and resonance fluorescence (light gray lines). Even though the low-energy fluorescence from $1s^{-1}2s^{-1}$,$2s^{-2}$, and $1s^{-1}2p^{-1}_{0,\pm1}$ and photoionization and Auger processes  to Ne$^{3+}$ are not shown, they are included in the calculations.}
\end{figure*}

\begin{comment}
\begin{figure}
\resizebox{85mm}{!}{\includegraphics{Ne2PiMasterEq.png}}
\caption{\label{MasterEq} Time-dependent population of Ne\textsuperscript{+} and Ne\textsuperscript{2+} exposed to a 0.25-fs pulse. (b) Time-dependent population of 4 electronic configurations of Ne\textsuperscript{+}. (c) Time-dependent population of 14 electronic configurations of Ne\textsuperscript{2+}.}
\end{figure}
\end{comment}

\begin{figure*}
\resizebox{180mm}{!}{\includegraphics{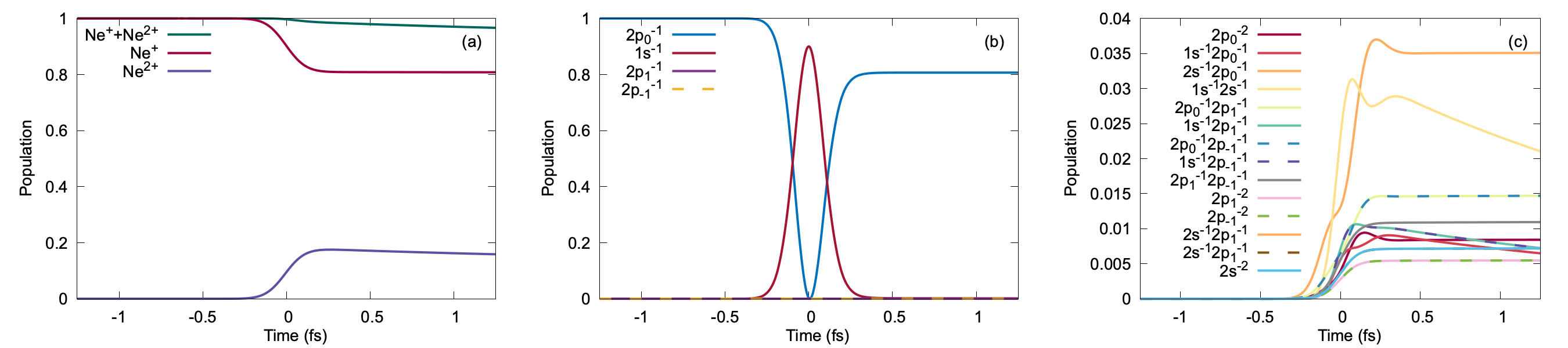}}
\caption{\label{MasterEqDyn} Time-dependent population of (a) Ne\textsuperscript{+} and Ne\textsuperscript{2+} (b) 4 electronic configurations of Ne\textsuperscript{+} and (c) 14 electronic configurations of Ne\textsuperscript{2+} computed for a 0.25-fs pulse with $Q=2\pi$. }
\end{figure*}

To examine how population dynamics in Ne$^{2+}$, produced from Ne$^{+}$ under an intense attosecond pulse, affect the scattering signal, we solve the time-dependent evolution of 18 electronic configurations using a density-matrix approach. The model includes four Ne$^{+}$ configurations,
\[
\ket{0}=\ket{2p^{-1}_{-1}},
\ket{1}=\ket{2p^{-1}_{0}},
\ket{2}=\ket{2p^{-1}_{1}},
\ket{3}=\ket{1s^{-1}},
\]
together with fourteen Ne$^{2+}$ configurations,
\[
\ket{4}=\ket{2p^{-2}_{0}},\;
\ket{5}=\ket{1s^{-1}2p^{-1}_{0}},\]
\[
\ket{6}=\ket{2s^{-1}2p^{-1}_{0}},\;
\ket{7}=\ket{1s^{-1}2s^{-1}},
\]
\[
\ket{8}=\ket{2p^{-1}_{0}2p^{-1}_{1}},\;
\ket{9}=\ket{1s^{-1}2p^{-1}_{1}},\]
\[
\ket{10}=\ket{2p^{-1}_{0}2p^{-1}_{-1}},\;
\ket{11}=\ket{1s^{-1}2p^{-1}_{-1}},
\]
\[
\ket{12}=\ket{2p^{-1}_{1}2p^{-1}_{-1}},\;
\ket{13}=\ket{2p^{-2}_{1}},\;
\ket{14}=\ket{2s^{-1}2p^{-1}_{1}},\]
\[
\ket{15}=\ket{2p^{-2}_{-1}},
\ket{16}=\ket{2s^{-1}2p^{-1}_{-1}},
\ket{17}=\ket{2s^{-2}}.
\]
The double-core-hole configuration $\ket{1s^{-2}}$ is excluded because it is not accessible via single-photon absorption at 849.8 eV. As shown in Fig.~\ref{Fig_18levels}, five pairs of states, one in Ne$^{+}$ and four in Ne$^{2+}$, are resonantly coupled by the driving field. The initial state $\ket{2p^{-1}_{0}}$ is resonantly coupled to $\ket{1s^{-1}}$. The density-matrix formulation enables us to track both the population and coherence within each Rabi-active pair, as well as the incoherent population transfer among the remaining eight levels.

The time evolution of the density-matrix elements 
$\rho_{a,b}(t)=\langle a|\hat{\rho}(t)|b\rangle$ is described by the general master equation \cite{Cavaletto_ResFluor_PRA, Kai_phay_Neplus,li2016coherence,lidar2019lecture,manzano2020short,Campaioli2024quantum}
\begin{align}
    \dot{\hat{\rho}}(t)
    = -i\bigl[\hat{H}(t),\hat{\rho}(t)\bigr]
      + \mathcal{L}\hat{\rho}(t)
      + \mathcal{D}\hat{\rho}(t),
\end{align}
where the first term generates the coherent, field-driven dynamics.  The Lindblad term $\mathcal{L}\hat{\rho}(t)$ accounts for all population-preserving incoherent transitions within the Ne$^{+}$/Ne$^{2+}$ 
manifold, including Auger decay, fluorescence, and photoionization channels whose final states lie within the 18-level model space.  The loss term $\mathcal{D}\hat{\rho}(t)$ collects all population-removing processes, including photoionization and Auger decay into Ne$^{3+}$ and higher charge states, which reduce the trace of $\hat{\rho}(t)$.

In this section, we solve a set of 23 coupled differential equations describing the populations of all 18 electronic configurations and the coherences between the five Rabi-active pairs. Explicit expressions for the equations of motion $\dot{\rho}_{a,b}(t)$ from the master equation are given below
\begin{widetext}
\begin{align}
\dot \rho_{1,1}(t) &= 
 \gamma^{F}_{3,1}\,\rho_{3,3}(t)-\gamma^{P}_{1,tot}(t)\,\rho_{1,1}(t)
+ i \bigl[\Omega_{3,1}(t)\rho_{1,3}(t) - c.c. \bigr],
\\[6pt]
\dot \rho_{3,3}(t) &= 
-\bigl[\gamma^{P}_{3,tot}(t)+\gamma^{A}_{3,tot}+\gamma^{F}_{3,tot}\bigr]\,\rho_{3,3}(t)
- i \bigl[\Omega_{3,1}(t)\rho_{1,3}(t) - c.c. \bigr],
\\[6pt]
\dot \rho_{1,3}(t) &= 
i\Delta_{3,1}\,\rho_{1,3}(t)
-\frac{1}{2}[\gamma^{P}_{1,tot}(t)+\gamma^{P}_{3,tot}(t)+\gamma^{A}_{3,tot}+\gamma^{F}_{3,tot}\bigr]\rho_{1,3}(t)
-i \Omega_{1,3}(t)\bigl[\rho_{3,3}(t)-\rho_{1,1}(t)\bigr],
\\[6pt]
\dot \rho_{4,4}(t) &= 
\gamma^{P}_{1,4}(t) \,\rho_{1,1}(t)
+\gamma^{A}_{3,4}\,\rho_{3,3}(t)+\gamma^{F}_{5,4}\,\rho_{5,5}(t)+\gamma^{F}_{6,4}\,\rho_{6,6}(t)
-\Gamma^{P}_{4}(t)\,\rho_{4,4}(t)
+ i \bigl[\Omega_{5,4}(t)\rho_{4,5}(t)-c.c.\bigr],
\\[6pt]
\dot \rho_{5,5}(t) &= 
\gamma^{P}_{3,5}(t)\,\rho_{3,3}(t)+\gamma_{7,5}^{F}\,\rho_{7,7}(t)
-\bigl[\Gamma_{5}^{P}(t)+\Gamma_{5}^{A}+\gamma_{5,tot}^{F}\bigr]\,\rho_{5,5}(t)
-i \bigl[\Omega_{5,4}(t)\rho_{4,5}(t)-c.c.\bigr],
\\[6pt]
\dot \rho_{4,5}(t) &= 
i\Delta_{5,4}\,\rho_{4,5}(t)
-\frac{1}{2}\bigl[\Gamma^{P}_{4}(t)+\Gamma_{5}^{P}(t)+\Gamma_{5}^{A}+\gamma_{5,tot}^{F}\bigr]\,\rho_{4,5}(t)
-i \Omega_{4,5}(t)\bigl[\rho_{5,5}(t)-\rho_{4,4}(t)\bigr],
\\[6pt]
\dot \rho_{6,6}(t) &= 
\gamma^{P}_{1,6}(t) \,\rho_{1,1}(t)
+\gamma^{A}_{3,6}\,\rho_{3,3}(t)+\gamma^{F}_{7,6}\,\rho_{7,7}(t)+\gamma^{F}_{17,6}\,\rho_{17,17}(t)
-\bigl[\Gamma^{P}_{6}(t)+\gamma^{F}_{6,tot}\bigr]\,\rho_{6,6}(t) \nonumber \\[6pt]
&+ i \bigl[\Omega_{7,6}(t)\rho_{6,7}(t)-c.c.\bigr],
\\[6pt]
\dot \rho_{7,7}(t) &= 
\gamma^{P}_{3,7}(t)\,\rho_{3,3}(t)
-\bigl[\Gamma_{7}^{P}(t)+\Gamma_{7}^{A}+\gamma_{7,tot}^{F}\bigr]\,\rho_{7,7}(t)
-i \bigl[\Omega_{7,6}(t)\rho_{6,7}(t)-c.c.\bigr],
\\[6pt]
\dot \rho_{6,7}(t) &= 
i\Delta_{7,6}\,\rho_{6,7}(t)
-\frac{1}{2}\bigl[\Gamma^{P}_{6}(t)+\Gamma_{7}^{P}(t)+\Gamma_{7}^{A}+\gamma_{6,tot}^{F}+\gamma_{7,tot}^{F}\bigr]\,\rho_{6,7}(t)
-i \Omega_{6,7}(t)\bigl[\rho_{7,7}(t)-\rho_{6,6}(t)\bigr],
\\[6pt]
\dot \rho_{8,8}(t) &= 
\gamma^{P}_{1,8}(t) \,\rho_{1,1}(t)+\gamma^{P}_{2,8}(t) \,\rho_{2,2}(t)
+\gamma^{A}_{3,8}\,\rho_{3,3}(t)+\gamma^{F}_{5,8}\,\rho_{5,5}(t)+\gamma^{F}_{6,8}\,\rho_{6,6}(t)+\gamma^{F}_{9,8}\,\rho_{9,9}(t)+\gamma^{F}_{14,8}\,\rho_{14,14}(t) \nonumber \\[6pt]
&-\Gamma^{P}_{8}(t)\,\rho_{8,8}(t)
+ i \bigl[\Omega_{9,8}(t)\rho_{8,9}(t)-c.c.\bigr],
\\[6pt]
\dot \rho_{9,9}(t) &= 
\gamma^{P}_{3,9}(t)\,\rho_{3,3}(t)+\gamma_{7,9}^{F}\,\rho_{7,7}(t)
-\bigl[\Gamma_{9}^{P}(t)+\Gamma_{9}^{A}+\gamma_{9,tot}^{F}\bigr]\,\rho_{9,9}(t)
-i \bigl[\Omega_{9,8}(t)\rho_{8,9}(t)-c.c.\bigr],
\\[6pt]
\dot \rho_{8,9}(t) &= 
i\Delta_{9,8}\,\rho_{8,9}(t)
-\frac{1}{2}\bigl[\Gamma^{P}_{8}(t)+\Gamma_{9}^{P}(t)+\Gamma_{9}^{A}+\gamma_{9,tot}^{F}\bigr]\,\rho_{8,9}(t)
-i \Omega_{8,9}(t)\bigl[\rho_{9,9}(t)-\rho_{8,8}(t)\bigr],
\\[10pt]
\dot \rho_{10,10}(t) &= 
\gamma^{P}_{1,10}(t) \,\rho_{1,1}(t)+\gamma^{P}_{0,10}(t) \,\rho_{0,0}(t)
+\gamma^{A}_{3,10}\,\rho_{3,3}(t)+\gamma^{F}_{5,10}\,\rho_{5,5}(t)+\gamma^{F}_{6,10}\,\rho_{6,6}(t)+\gamma^{F}_{11,10}\,\rho_{11,11}(t)  \nonumber \\[6pt]
&+\gamma^{F}_{16,10}\,\rho_{16,16}(t)-\Gamma^{P}_{10}(t)\,\rho_{10,10}(t)
+ i \bigl[\Omega_{11,10}(t)\rho_{10,11}(t)-c.c.\bigr],
\\[6pt]
\dot \rho_{11,11}(t) &= 
\gamma^{P}_{3,11}(t)\,\rho_{3,3}(t)+\gamma_{7,11}^{F}\,\rho_{7,7}(t)
-\bigl[\Gamma_{11}^{P}(t)+\Gamma_{11}^{A}+\gamma_{11,tot}^{F}\bigr]\,\rho_{11,11}(t)
-i \bigl[\Omega_{11,10}(t)\rho_{10,11}(t)-c.c.\bigr],
\\[6pt]
\dot \rho_{10,11}(t) &= 
i\Delta_{11,10}\,\rho_{10,11}(t)
-\frac{1}{2}\bigl[\Gamma^{P}_{10}(t)+\Gamma_{11}^{P}(t)+\Gamma_{11}^{A}+\gamma_{11,tot}^{F}\bigr]\,\rho_{10,11}(t)
-i \Omega_{10,11}(t)\bigl[\rho_{11,11}(t)-\rho_{10,10}(t)\bigr],
\\[6pt]
\dot \rho_{0,0}(t) &=  \gamma^{F}_{3,0}\,\rho_{3,3}(t)
-\gamma^{P}_{0,tot}(t)\,\rho_{0,0}(t),
\\[6pt]
\dot \rho_{2,2}(t) &=  \gamma^{F}_{3,2}\,\rho_{3,3}(t)
-\gamma^{P}_{2,tot}(t)\,\rho_{2,2}(t),
\\[6pt]
\dot \rho_{12,12}(t) &=  
\gamma^{P}_{0,12}(t) \,\rho_{0,0}(t)+\gamma^{P}_{2,12}(t) \,\rho_{2,2}(t)+\gamma^{A}_{3,12}\,\rho_{3,3}(t)+\gamma^{F}_{9,12}\,\rho_{9,9}(t)+\gamma^{F}_{11,12}\,\rho_{11,11}(t)+\gamma^{F}_{14,12}\,\rho_{14,14}(t)
\nonumber 
\\[6pt]
&+\gamma^{F}_{16,12}\,\rho_{16,16}(t)-\Gamma^{P}_{12}(t)\,\rho_{12,12}(t),
\\[6pt]
\dot \rho_{13,13}(t) &=  
\gamma^{P}_{0,13}(t) \,\rho_{0,0}(t)+\gamma^{A}_{3,13}\,\rho_{3,3}(t)+\gamma^{F}_{9,13}\,\rho_{9,9}(t)+\gamma^{F}_{14,13}\,\rho_{14,14}(t)
-\Gamma^{P}_{13}(t)\,\rho_{13,13}(t),
\\[6pt]
\dot \rho_{14,14}(t) &=  
\gamma^{P}_{0,14}(t) \,\rho_{0,0}(t)+\gamma^{A}_{3,14}\,\rho_{3,3}(t)+\gamma^{F}_{7,14}\,\rho_{7,7}(t)+\gamma^{F}_{17,14}\,\rho_{17,17}(t)
-\bigl[\Gamma^{P}_{14}(t)+\gamma^{F}_{14,tot}\bigr]\,\rho_{14,14}(t),
\\[6pt]
\dot \rho_{15,15}(t) &=  
\gamma^{P}_{2,15}(t) \,\rho_{2,2}(t)+\gamma^{A}_{3,15}\,\rho_{3,3}(t)+\gamma^{F}_{11,15}\,\rho_{11,11}(t)+\gamma^{F}_{16,15}\,\rho_{16,16}(t)
-\Gamma^{P}_{15}(t)\,\rho_{15,15}(t),
\\[6pt]
\dot \rho_{16,16}(t) &=  
\gamma^{P}_{2,16}(t) \,\rho_{2,2}(t)+\gamma^{A}_{3,16}\,\rho_{3,3}(t)(t)+\gamma^{F}_{7,16}\,\rho_{7,7}(t)+\gamma^{F}_{17,16}\,\rho_{17,17}(t)
-\bigl[\Gamma^{P}_{16}(t)+\gamma^{F}_{16,tot}\bigr]\,\rho_{16,16}(t),
\\[6pt]
\dot \rho_{17,17}(t) &=  
\gamma^{A}_{3,17}\,\rho_{3,3}(t)
-\bigl[\Gamma^{P}_{17}(t)+\gamma^{F}_{17,tot}\bigr]\,\rho_{17,17}(t),
\end{align}
\end{widetext}
where $\Omega_{a,b}(t)= E(t)d_{a,b}$, $d_{a,b}$ and $\Delta_{a,b} = E_a-E_b$ are Rabi frequency, dipole moment and energy difference between levels $a$ and $b$. The rates $\gamma^{X}_{a,b}$ ($X=\mathrm{A,F,P}$) describe population-preserving Auger (A), fluorescence (F), and photoionization (P) transitions from state $a$ to state $b$ within the 18-level manifold. The quantity $\gamma^{X}_{a,tot} = \sum_{b}\gamma^{X}_{a,b}$ is the total rate for  process $X$ out of state $a$. The rates $\Gamma^{X}_{a}$ denote the photoionization and Auger channels that remove population from state $a$ and populate Ne$^{3+}$ outside the model space. We note that $\Omega_{a,b}(t)= E(t)d_{a,b}= (\Omega_{b,a}(t))^{*}$ and $\rho_{a,b}(t)=(\rho_{b,a}(t))^{*}$ are hermitian.
All the rates, cross section, dipole moments and energies are computed using our previously developed Hartree-Fock-Slater electronic structure code \cite{Ho-2014-PRL,Ho-2017-JPB}.

Figure~\ref{MasterEqDyn} shows the population dynamics of all 18 electronic configurations during a 0.25-fs pulse. As expected for a $2\pi$ pulse, Ne$^{+}$ undergoes a full Rabi cycle. Various Ne$^{2+}$ states are populated incoherently through photoionization and inner-shell decay of Ne$^{+}$ during the pulse. Despite being more than 6 eV detuned from their respective resonances, the four Ne$^{2+}$ Rabi-active pairs nonetheless undergo some coherent population exchange. A fraction of Ne$^{2+}$ is further photoionized and subsequently decays via Auger processes, producing Ne$^{3+}$ and higher charge states that lie outside our model space.

At the peak of the 0.25-fs pulse ($t = 0$), the populations of Ne$^{+}$ and Ne$^{2+}$ are 0.9020 and 0.0911, respectively; these values set an upper bound on the possible scattering contribution from Ne$^{2+}$. A similar calculation for a $Q = 1.5\pi$ pulse yields peak populations of 0.9395 (Ne$^{+}$) and 0.0594 (Ne$^{2+}$).

We note that the present calculations include only the unscattered electronic states for a single atom. A complete determination of the total coherent scattering signal, including Ne$^{+}$–Ne$^{2+}$ and Ne$^{2+}$–Ne$^{2+}$ interference, would require explicitly incorporating all one-photon–scattered states for two atoms into the density matrix. This extension is beyond the scope of the current work and is unnecessary for our purposes, as the scattering contributions from different ionic pairs produce only a near constant angular background compared with the dominant scattering from Ne$^{+}$–Ne$^{+}$ pairs.

\section{Effects of Intensity on the Angular Distribution of Resonance Fluorescence} \label{App_RFtrendintensity}

\begin{figure}
\vspace{15pt}
\resizebox{80mm}{!}{\includegraphics{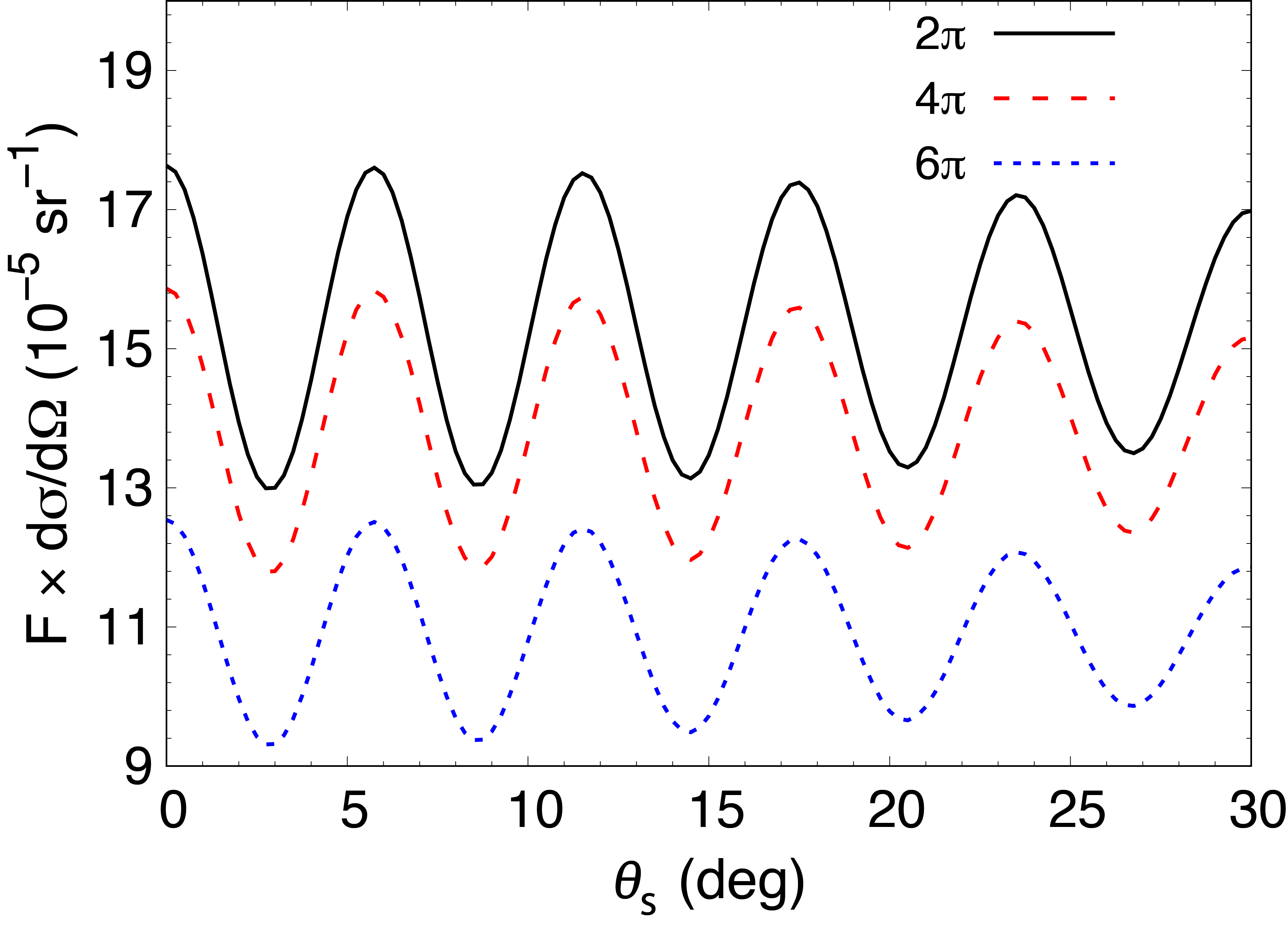}}
\caption{\label{Fig_RFtrendintensity} Angular dependence of resonance fluorescence from two atoms for an incident pulse of duration $0.25$ fs for increasing intensities. The initial state of each atom is $\ket{1}$. The other parameters are the same as Fig.~\ref{Fig_2Pi_signal_initialstate2pz} }.
\end{figure}

Here we examine how the interference in resonance fluorescence changes with increasing pulse area, which corresponds to increasing intensity for a given pulse duration. The calculated results shown in Fig.~\ref{Fig_RFtrendintensity} are computed for three different pulse areas of $Q = 2\pi$, $4\pi$, and $6\pi$ which correspond to intensities of about $2.1\times 10^{18}$ W/cm\textsuperscript{2}, $8.3\times 10^{18}$ W/cm\textsuperscript{2}, and $1.9\times 10^{19}$ W/cm\textsuperscript{2}, respectively. The calculations demonstrate that the fringe contrast in the angular distribution decreases with increasing intensity, and therefore the interference is progressively reduced. This trend is consistent with earlier studies of resonance fluorescence under strong driving using monochromatic optical fields~\cite{Richter1991_RFinterference,agarwal2002_Iorder_vs_IIorder,Keitel2007_strongfield_interf}. The reduction of first-order interference at high intensities has important implications for resonant x-ray scattering in single-particle imaging, where multiple Rabi oscillations can significantly diminish coherent contrast.

  % End of color red

\bibliography{References.bib}

\end{document}